\documentclass[12pt]{article}
\usepackage{epsfig,amssymb,amsmath,psfrag}
\usepackage{bm}

\catcode`\@=11

\def \bb  {\begin {thebibliography} }
\def \eb  {\end{thebibliography}}
\def \lab #1 {\label{#1}}

\newcommand\re[1]{(\ref{#1})}

\def \qqqquad {\qquad\qquad}
\def \matrix #1 {\left(\begin{array}{cc} #1 \end{array}\right)}

\def \tr {\mathop{\rm tr}\nolimits}

\def \e  {\mathop{\rm e}\nolimits}
\newcommand\lr[1]{{\left({#1}\right)}}
\newcommand \widebar [1] {\overline{#1}}

\newcommand \vev [1] {\langle{#1}\rangle}

\newcommand \ket [1] {|{#1}\rangle}

\newcommand{\ft}[2]{{\textstyle\frac{#1}{#2}}}

\def \thesection {\arabic{section}.}

\def\numberbysection{\@addtoreset{equation}{section}
                     \def\theequation{\thesection\arabic{equation}}}

\numberbysection

\begin{document}

\begin{titlepage}
\begin{flushright}
\begin{tabular}{l}
IPhT--T13/145 
\end{tabular}
\end{flushright}

\vskip3cm

\centerline{\large \bf Dual conformal symmetry on the light-cone}

\vspace{1cm}

\centerline{\sc  S.\'E. Derkachov$^{a}$,
                G.P. Korchemsky$^b$, A.N. Manashov$^{c,d}$}

\vspace{10mm}

\centerline{\it  $^a$ St. Petersburg Department of Steklov Mathematical Institute, }
\centerline{\it  Fontanka 27, 191023 St. Petersburg, Russia}

\vspace{3mm}

\centerline{\it $^b$Institut de Physique Th\'eorique\,\footnote{Unit\'e de Recherche
Associ\'ee au CNRS URA 2306}, CEA Saclay, }\centerline{\it 91191 Gif-sur-Yvette Cedex,
France}

\vspace{3mm}

\centerline{\it $^c$Department of Theoretical Physics,  St.-Petersburg State University}

\centerline{\it 199034, St.-Petersburg, Russia}

\vspace{3mm}

\centerline{\it $^d $Institut f{\"u}r Theoretische Physik, Universit{\"a}t Regensburg}
\centerline{\it D-93040 Regensburg, Germany}

\def\thefootnote{\fnsymbol{footnote}}%
\vspace{1cm}

\centerline{\bf Abstract}

\vspace{5mm}

We study the properties of conformal operators in the $SL(2)$ sector of planar $\mathcal
N=4$ SYM and its supersymmetric $SL(2|2)$ extension. The correlation functions of these
operators and their form factors with respect to asymptotic on-shell states are determined
in the appropriate limit by two different polynomials which can be identified as
eigenstates of the dilatation operator in the coordinate and momentum representations,
respectively. We argue that, in virtue of integrability of the dilatation operator, the
two polynomials satisfy a duality relation -- they are proportional to each other upon an
appropriate identification of momenta and coordinates. Combined with the conventional
$\mathcal{N}=4$ superconformal symmetry, this leads to the dual  superconformal symmetry
of the dilatation operator. We demonstrate that this symmetry is powerful enough to fix
the eigenspectrum of the dilatation operator to the lowest order in the coupling. We use
the relation between the one-loop dilatation operator and Heisenberg spin chain to show
that, to lowest order in the coupling, the dual symmetry is generated by the Baxter
$Q-$operator in the limit of large spectral parameter.

\end{titlepage}

\setcounter{footnote} 0

\thispagestyle{empty}

\newpage

\pagestyle{plain} \setcounter{page} 1

\section{Introduction}

In this paper, we discuss the relation between dual superconformal symmetry in planar
$\mathcal{N}=4$ SYM \cite{Drummond:2008vq} and integrability of dilatation operator in the
same theory (for a review, see \cite{Beisert:2010jr}). At present, the dual symmetry is
best understood for scattering amplitudes through their duality to light-like polygon (super)
Wilson loops \cite{Alday:2007hr,Drummond:2007aua,Brandhuber:2007yx,Mason:2010yk,CaronHuot:2010ek,Belitsky:2011zm} and to
correlation functions in the light-cone limit \cite{Eden:2011yp,Eden:2011ku,Adamo:2011dq}. In a generic
Yang-Mills theory, these objects depend on two different sets of variables (on-shell
momenta of scattering particles versus coordinates of local operators in Minkowski
space-time)  and are not related to each other in a simple way. The very fact that such a
relation exists in planar $\mathcal{N}=4$ SYM immediately leads to an enhancement of the
symmetry -- the conventional  $\mathcal{N}=4$ superconformal symmetry of Wilson loops and
correlation functions combined with the duality relation imply the dual superconformal
symmetry of the scattering amplitudes.

The manifestation of the dual conformal symmetry can be also found in gauge theories with
less supersymmetry including QCD. In particular, the dual symmetry has first emerged as
the property of a particular class of scalar four-dimensional Feynman integrals
\cite{Broadhurst:1993ib,Drummond:2006rz}. It was also identified as the hidden symmetry of
the BFKL equation \cite{Lipatov:1998as} and its generalisations
\cite{Gomez:2009bx,Prygarin:2009zz} and of the evolution equations governing the scale
dependence of distribution amplitudes in QCD \cite{Braun:1999te}.

The dual superconformal symmetry  acts naturally on the space of dual coordinates
$(x_i,\theta_i)$. They are related to the (super)momenta of scattering on-shell states
$(p_i,\eta_i)$ as \cite{Drummond:2008vq}
\begin{align}\label{dual-coor}
& p_i^{\alpha\dot\alpha}=(x_i-x_{i+1})^{\alpha\dot\alpha}\,,\qquad
 \eta_i^A \lambda_i^\alpha = (\theta_i-\theta_{i+1})^{A\,\alpha}\,,
\end{align}
where $p_i^{\alpha\dot\alpha}= \lambda_i^\alpha \tilde\lambda_i^{\dot\alpha}$ are
light-like momenta of particles in the spinor-helicity notations (with
$\alpha,\dot\alpha=1,2$) and Grassmann variables $\eta_i^A$ (with $A=1,\dots,4$) serve to
combine all asymptotic states into a single  $\mathcal{N}=4$ on-shell superstate
\cite{Nair:1988bq}. The dual symmetry is the exact symmetry of the scattering amplitudes
in planar $\mathcal{N}=4$ SYM at tree level only. At loop level, it is believed that the
scattering amplitudes in planar $\mathcal{N}=4$ SYM also respect the dual symmetry for
arbitrary coupling, albeit in its anomalous form
\cite{Drummond:2007au,Korchemsky:2009hm,CaronHuot:2011kk,Bullimore:2011kg}.

In  AdS/CFT description of the scattering amplitudes \cite{Alday:2007hr}, the dual
conformal symmetry arises at strong coupling from the symmetry of sigma-model on
AdS${}_5\times$S${}^5$ background under the combined bosonic and fermonic T--duality
\cite{Kallosh:1998ji,Berkovits:2008ic,Beisert:2008iq}. Indeed, this sigma model is
integrable and it possesses a lot of symmetries generated by the conserved charges
\cite{Mandal:2002fs,Bena:2003wd}. The latter have been thoroughly studied in application
to the energies of stringy excitations \cite{Beisert:2010jr}, 
or equivalently the eigenvalues of the dilatation operator in planar $\mathcal{N}=4$ SYM. The
AdS/CFT correspondence suggests that, despite the fact that the scattering amplitudes and
dilatation operator have different meaning in planar $\mathcal{N}=4$ SYM, they should have
the same symmetries at strong coupling related to those of sigma-model on
AdS${}_5\times$S${}^5$. We can therefore ask what does integrability of the dilatation
operator imply for the properties of the scattering amplitudes (and the $S-$matrix in
general) and, vice versa, what is the manifestation of the dual conformal symmetry for the
dilatation operator in planar $\mathcal{N}=4$ SYM?

To address this question, we extend the dual symmetry to the so-called light-ray operators
in $\mathcal{N}=4$ SYM
\begin{align}\label{O-def}
\mathbb{O}(\bm{z}) = \tr\left[ Z(nz_1) \ldots Z(n z_L)\right].
\end{align}
These are nonlocal single-trace operators built from $L$ copies of a complex scalar field
$Z(x) = Z^a (x) T^a$, with $T^a$ being generators of the fundamental representation of the
$SU(N_c)$ gauge group. All scalar fields in \re{O-def} are located along the same light
ray defined by the light-like vector $n^\mu$ and variables $\bm{z}=(z_1,\ldots,z_L)$
denote the set of (real valued) light-cone coordinates. It is tacitly assumed that the
gauge invariance of \re{O-def} is restored by inserting the path ordered exponentials
$P\exp\big(ig\int_{ n z_i}^{n z_{i+1}} dx\cdot A(x)\big)$ between the adjacent scalar
fields  on the right-hand side of \re{O-def}. Such factors can be avoided by choosing the
gauge $(n\cdot A(x))=0$.

The light-ray operators allow us to define two different functions that we shall denote as
$\Phi_\alpha({\bm z})$ and $\Psi _\alpha ({\bm p})$. The former depends on the
light-cone coordinates of scalar fields and it is closely related to the operator product
expansion of $\mathbb{O}(\bm{z})$. Namely, expansion of the light-ray operator \re{O-def}
around $z_i=0$ produces an infinite set of local operators which mix under renormalization
and form a closed $SL(2)$ sector in $\mathcal{N}=4$ SYM. Diagonalizing the corresponding
mixing matrix, we can construct the conformal operators $ \mathcal{O}_{\alpha}(0)$ having
an autonomous scale dependence. Then, the expansion of the light-ray operators over the
basis of conformal operators takes the form
\begin{align}\label{acd}
 \mathbb{O}({\bm z}) = \sum_{\alpha}  \Phi_\alpha({\bm z})   {\mathcal O}_\alpha(0)\,,
\end{align}
where the coefficient functions $\Phi_\alpha({\bm z})$ are homogenous polynomials
depending on light-cone coordinates of scalar fields (as well as on the coupling
constant). The explicit form of $\Phi_\alpha({\bm z})$ can be found by diagonalizing the
dilatation operator in the $SL(2)$ sector.

The second function, $\Psi _\alpha ({\bm p})$, is the form factor defined as the matrix
element of the conformal operator ${\mathcal O}_\alpha(0)$
\begin{align}\label{bcd}
\vev{0|{\mathcal O}_\alpha(0) |P_1,\dots,P_L} \sim  \widebar \Psi _\alpha ({\bm p}) \,,\qquad
P_i^\mu = p_i \,\bar n^\mu\,,
\end{align}
where complex conjugation $\widebar \Psi _\alpha ({\bm p}) \equiv \lr{ \Psi _\alpha ({\bm
p})}^* $ is introduced for the later convenience. Here the asymptotic state
$\ket{P_1,\dots,P_L}$ consists of $L$ massless particles carrying the momenta aligned
along the same light-cone direction  $\bar n^\mu$ (with $(n\bar n)\neq 0$ and ${\bm
p}=(p_1,\dots,p_L)$ being the corresponding light-cone components). The reason for such
choice of particle momenta is motivated by the previous studies of analogous matrix elements  in QCD. The matrix
elements of the form \re{bcd} naturally appear in QCD description of hadrons as bound
states of partons (quarks and gluons). In virtue of asymptotic freedom, the interaction
between partons becomes weak at high energy. Therefore, when boosted into an infinite
momentum frame, the hadron behaves as a collection of noninteracting partons moving along
the same light-cone direction with the momenta $P_i^\mu = p_i \,\bar n^\mu$. Then, the
function $\Psi _\alpha ({\bm p})$ defines the projection of the composite state ${\mathcal
O}_\alpha(0)\ket{0}$ onto one of its Fock components $\ket{P_1,\dots,P_L}$ and has the
meaning of light-cone distribution amplitude (for a review, see
\cite{Lepage:1980fj,Chernyak:1983ej,Braun:2003rp}).

As follows from their definition \re{acd} and \re{bcd}, the functions $ \Phi_\alpha({\bm
z})$ and $\Psi _\alpha ({\bm p})$ have a different interpretation and  should
be independent on each other.  Nevertheless, previous studies of
three-particle (baryon) distribution amplitudes in QCD revealed \cite{Braun:1999te} that
the two functions are proportional to each other, to one-loop order at least, upon
identification of light-cone momenta and coordinates, $p_i=z_i-z_{i+1}$. In the present
paper, we generalise this relation in planar $\mathcal N=4$ SYM to the states of arbitrary
length $L$,
\begin{align}\label{ann}
 \Psi _\alpha ({\bm p}) = \xi_\alpha \Phi_\alpha({\bm z}) \,,\qqqquad p_i=z_i-z_{i+1}\,,
\end{align}
with $z_{L+1}=z_1$. Notice that the proportionality factor $ \xi_\alpha$ only depends on
the quantum numbers of the conformal primary operator and on the coupling constant, but
not on the dual coordinates. The relation $p_i=z_i-z_{i+1}$ is very similar to the first
relation in \re{dual-coor}. In fact, the two relations are equivalent once we restrict the
dual coordinates $x_i^\mu$ to be aligned along the same light-cone direction
\begin{align}
P_i^\mu = \bar n^\mu p_i\,,\qquad x_i^\mu  = \bar n^\mu z_i \,.
\end{align}
The relation \re{ann} is similar to the duality relation between
scattering amplitudes and light-like Wilson loops mentioned above but this time it establishes
the correspondence between the coefficient functions and form factors of conformal primary operators.

To lowest order in the coupling, the duality relation \re{ann} follows from integrability
of the $SL(2)$ dilatation operator. More precisely, to one-loop order  the functions
$\Phi_\alpha({\bm z})$ and $\Psi _\alpha ({\bm p})$ coincide  with eigenstates of the
$SL(2)$ Heisenberg spin chain in the coordinate and momentum representations,
respectively, and their symmetry properties can be studied with a help of the Baxter
$Q-$operator. This operator was first introduced by Baxter in solving the 8-vertex model
\cite{Baxter:1972hz} and has proven to be a very powerful tool in solving a variety of
integrable models \cite{Bazhanov:1989nc,Gaudin:1992ci,Bazhanov:1996dr}. In the case of the
$SL(2)$ Heisenberg spin chain, the Baxter operator $\mathbb{Q}(u)$ depends on an arbitrary
complex parameter $u$ and satisfies the defining relations summarised below in Sect.~4.1.
Its explicit construction was carried out in  Ref.~\cite{Derkachov:1999pz}.

The Baxter $Q-$operator is the generating function of integrals of motions of the $SL(2)$
spin chain. In particular, the one-loop dilatation operator in $\mathcal N=4$ SYM can be
obtained from expansion of $\mathbb{Q}(u) $ around $u=\pm i/2$.  As a consequence, to
one-loop order, the functions $\Phi_\alpha({\bm z})$ have to diagonalise the operator
$\mathbb{Q}(u)$ for any $u$. The dual symmetry arises when we examine the action of the
Baxter $Q-$operator on $\Phi_\alpha({\bm z})$ for large values of the spectral parameter,
$u\to\infty$, \cite{Derkachov:1999pz}
\begin{align}\label{Q-gen}
\mathbb{Q}(u) \Phi_\alpha({\bm z}) \sim  \Psi _\alpha ({\bm p})+O(1/u)\,,\qquad p_i=z_i-z_{i+1}\,.
\end{align}
Since $\Phi_\alpha({\bm z})$ diagonalises the Baxter $Q-$operator, it follows from this
relation that $\Phi_\alpha({\bm z})$ has to be proportional to $ \Psi _\alpha ({\bm p})$
thus leading to \re{ann}.  In this manner, the duality relation \re{ann} is generated, to
the lowest order in the coupling,  by the leading term in the asymptotic expansion of
Baxter $Q-$operator at infinity.

According to \re{Q-gen}, the Baxter operator automatically generates the transition to the
dual coordinates, $p_i=z_i-z_{i+1}$, thus equating to zero the total momentum of $ \Psi
_\alpha ({\bm p})$. This means that, in distinction with the conventional conformal
symmetry, the dual conformal symmetry is only present  for the vanishing total momentum,
$\sum_i p_i=0$. The same property has been previously observed in the analysis of
scattering amplitudes and form factors in planar $\mathcal{N}=4$ SYM. For the scattering
amplitudes, the condition $\sum_i P_i^\mu=0$ is automatically satisfied. For the form
factors, $F_O=\int d^4 x
\e^{ix\cdot q}\vev{0|O(x)|P_1,\dots,P_n}$, the total momenta equals the momentum
transferred, $\sum_i P_i^\mu=q^\mu$. At weak coupling, the explicit calculation of form
factors showed  \cite{Brandhuber:2010ad} that the dual conformal symmetry is only present
for the vanishing momentum transferred, $q^\mu=0$. At strong coupling, the same result
follows from a dual description of the form factor   \cite{Alday:2007he} in terms of
minimal area attached to infinitely periodic zig-zag light-like contour located at the
boundary of the AdS${}_5$ and built from light-like momenta $P_i^\mu$. The dual conformal
symmetry of the form factor is broken for $q^\mu\neq 0$ because the above mentioned
kinematical configuration is not stable under conformal transformations.

The duality relation \re{ann} can be extended to a larger class of supersymmetric
light-ray operators. These operators are obtained from \re{O-def} by replacing the scalar
field ${Z}(z_i n)$ with gaugino and gauge strength fields in $\mathcal N=4$ SYM. To deal
with such operators it is convenient to employ the light-cone superspace formalism
\cite{Mandelstam:1982cb,Brink:1982pd}. It allows us to combine various components of
fields into a single light-cone superfield $\mathcal{Z}(z_i n,\theta_i)$ (with
$i=1,\dots,L$) and use it to construct the corresponding supersymmetric light-ray operator
\cite{Belitsky:2004sc,Belitsky:2005gr}. Going through the same steps as before, we can
define supersymmetric extension of the coefficient functions $\Phi_\alpha({\bm Z})$ and
form factors $ \Psi _\alpha ({\bm P})$ depending, respectively, on the set of $L$
light-cone supercoordinates ${\bm Z}=\{z_i,\theta_i\}$ and conjugated supermomenta ${\bm
P}=\{p_i,\vartheta_i\}$. The functions entering the duality relation \re{ann} are
the lowest components in the expansion of $\Phi_\alpha({\bm Z})$ and  $ \Psi _\alpha ({\bm
P})$ in powers of Grassmann $\theta-$variables. We show in this paper that the duality
relation also holds for the remaining components
\begin{align}\label{super-ann}
\Psi _\alpha ({\bm P}) = \xi_\alpha \Phi_\alpha({\bm Z})\,,\qquad p_i=z_i-z_{i+1}\,, \qquad \vartheta_i^A=\theta_i^A-\theta_{i+1}^A\,.
\end{align}
As compared with the general form of duality transformation \re{dual-coor}, the last two
relations in \re{super-ann} correspond to the collinear limit  $x_i^{\alpha\dot\alpha} =z_i 
n^{\alpha\dot\alpha} $ and $\theta_{i}^{A\,\alpha} = \theta_i^A\lambda ^\alpha$ (with
$n^{\alpha\dot\alpha}=\lambda ^\alpha\tilde\lambda^{\dot\alpha}$). As before, to the lowest
order in the coupling, the duality relation \re{super-ann} is generated by the Baxter
$Q-$operator for supersymmetric generalisation of the $SL(2)$ Heisenberg spin
chain \cite{Belitsky:2006cp,Bazhanov:2008yc}.

The paper is organised as follows. In Sect.~2 we describe the properties of light-ray
operators in $\mathcal N=4$ SYM and formulate the duality relation \re{ann}. In Sect.~3 we
verify this relation at one loop by diagonalising the dilatation operator in the $SL(2)$
sector. In Sect.~4 we explain the origin of the dual conformal symmetry of the one-loop
$SL(2)$ dilatation operator and demonstrate that it is generated by the leading term in
the asymptotic expansion of the Baxter $Q-$operator for large spectral parameter. We also
argue that the dual symmetry is powerful enough to uniquely fix the eigenstates of the
one-loop dilatation operator. In Sect.~5 we discuss supersymmetric extension of the
duality relation for a larger class of light-ray operators involving various components of
gaugino and gauge fields in $\mathcal{N}=4$ SYM. Concluding remarks are presented in
Sect.~6.

\section{Light-ray operators}

According to definition \re{O-def}, the light-ray operator $\mathbb{O}(\bm{z})$ is given
by the product of scalar fields located on the same light ray. Its expansion in powers of
$\bm{z}=(z_1,\ldots,z_L)$  produces an infinite set of local single-trace operators
\begin{align}\notag\label{start}
& \mathbb{O}(\bm{z}) = \sum_{\bm{k}} {z_1^{k_1}} \dots {z_L^{k_L}} \,  O_{\bm{k}}(0) \,,
 \qquad
 \\
 & O_{\bm{k}}(0) =  \tr\left[ {D_+^{k_1} \over k_1!} Z(0) \ldots {D_+^{k_L} \over k_L!} Z(0)\right] ,
\end{align}
where the sum runs over nonnegative integers $\bm{k}=(k_1,\ldots,k_L)$ and   $D_+ = (n D)$
stands for the light-cone component of the covariant derivative. 

It is tacitly assumed that the
operators $O_{\bm{k}}(0)$ are renormalised in a particular scheme (say minimal subtraction scheme)
and depend on the renormalisation scale. The operators
$O_{\bm{k}}(0)$ mix with each other under the change of this scale but
we can diagonalize their mixing
matrix and define the operators $ {\mathcal O}_\alpha(0)$ having an autonomous scale
dependence. They are given by a linear combination of the basis operators $O_{\bm k}(0)$
\begin{align}\label{auto}
  {\mathcal O}_\alpha(0) =  \sum_{\bm{k}} \widebar c_{\bm{k},\alpha}(g^2)\,{O}_{\bm k}(0) \,,
\end{align}
with the expansion coefficients depending on 't Hooft coupling constant $g^2=g_{\rm YM}^2
N/(8\pi^2)$. The coefficients $\bar c_{\bm{k},\alpha}=\lr{c_{\bm{k},\alpha}}^*$ coincide with the eigenstates of the
all-loop mixing matrix in planar $\mathcal{N}=4$ SYM and are labelled by the index
$\alpha$ (to be specified below).

\subsection{Operator product expansion}

Let us introduce the following polynomial
\begin{align}\label{Phi-def}
\Psi_\alpha ({\bm   p}) = \sum_{{\bm   k}} c_{{\bm k},\alpha}(g^2)\,{ p_1^{k_1}\over k_1!} \ldots {p_L^{k_L}\over k_L!}\,.
\end{align}
It involves the same expansion coefficients as \re{auto} and depends on the set of
auxiliary variables ${\bm p}=(p_1,\ldots,p_L)$. The polynomial \re{Phi-def} defines the
symbol of the differential operator $\widebar\Psi_\alpha ({\bm \partial_z})$ which
projects the light-ray operator $\mathbb{O}({\bm z})$ onto local operator $ {\mathcal
O}_\alpha(0)$. Namely, the operator \re{auto} is obtained from the light-ray
operator \re{O-def} by substituting $p_i\to \partial_{z_i}$ on the right-hand side of
\re{Phi-def} and  applying the resulting differential operator $\widebar\Psi_\alpha
({\bm \partial_z})$ to both sides of \re{O-def}
\begin{align}\label{O-hat}
  {\mathcal O}_\alpha(0) = \widebar\Psi_\alpha ({\bm \partial_z})\mathbb{O}({\bm z}) \big|_{{\bm z}=0}\,.
\end{align}
The $p_i-$variables in \re{Phi-def} are conjugated to light-cone coordinates $z_i$ of
scalar fields and have the meaning of the light-cone components of the momenta carried by
scalar particles.

Inverting  \re{auto},  we can expand ${O}_{\bm k}(0)$ over the basis of conformal
operators $ {\mathcal O}_\alpha(0)$ and rewrite the first relation in \re{start} as
\begin{align}\label{imp}
\mathbb{O}({\bm z}) = \sum_{\alpha}  \Phi_\alpha({\bm z})   {\mathcal O}_\alpha(0)\,,
\end{align}
where $\Phi_\alpha({\bm z})$ are (homogenous) polynomials depending on the light-cone
coordinates of  scalar fields. The main advantage of \re{imp} as compared with the
first relation in \re{start} is that each term on the right-hand side of \re{imp} has a definite scaling
dimension.

The two polynomials entering the right-hand side of \re{O-hat} and \re{imp} carry a
different information: $\Psi_\alpha ({\bm p})$ fixes the form of the local operator
\re{O-hat}, whereas  $\Phi_\alpha({\bm z}) $ determines its contribution to the  
operator expansion  \re{imp}. Since the light-ray operator \re{O-def} is invariant under
the cyclic shift of scalar fields inside the trace, the polynomials should be cyclically
invariant functions of their arguments.~\footnote{This property is ultimately related to
the fact that the light-ray operator \re{O-def} is built from the same complex field. If
the operator \re{O-def} involved different fields, as it happens in QCD, the above
condition should be relaxed.} Together with \re{Phi-def} this implies that $c_{{\bm
k},\alpha}(g^2)$ should be invariant under the cyclic shift of indices, $k_i\to k_{i+1}$.

The polynomials $\Psi_\alpha ({\bm p})$ and $\Phi_\alpha({\bm z})$ are not independent on
each other. Substituting \re{imp} into the right-hand side of \re{O-hat} and comparing the
coefficients in front of $ {\mathcal O}_\alpha(0)$, we find that the polynomials have to satisfy the
orthogonality condition
\begin{align}\label{orth}
 \widebar\Psi_\alpha ({\bm \partial_z})\Phi_\beta({\bm z}) \big|_{{\bm z}=0} = \delta_{\alpha\beta}\,.
\end{align}
In the similar manner, substitution of \re{O-hat} into \re{imp} yields the completeness
condition
\begin{align}\label{comp}
 \sum_\alpha \Phi_\alpha({\bm z})\widebar\Psi_\alpha ({\bm p}) = \frac1{L}\sum_{i=1}^L \exp\lr{p_1 z_i + \ldots+ p_L z_{i+L-1 }} \,,
\end{align}
where $z_{i+L}\equiv z_i$ and expressions on both sides of \re{comp} are invariant under cyclic
shifts of ${\bm z}$ and ${\bm p}$. Note that the relations \re{orth} and \re{comp} should
hold for arbitrary  coupling constant, independently on the choice of the renormalisation scheme. 
 
\subsection{Conformal symmetry}

Let us specify the quantum numbers of conformal operators $ {\mathcal O}_\alpha(0)$.
According to \re{auto}, these operators are given by a linear combination of the basis operators
${O}_{\bm k}(0)$  built from $L$ scalar fields and carrying the Lorentz
spin $S=\sum_i k_i$ equal to the total number of covariant derivatives. In addition, the
operators $ {\mathcal O}_\alpha(0)$ have a definite scaling dimension $\Delta_{S,\alpha}$
\begin{align}\label{Delta}
\Delta_{S,\alpha} = L+S +  \gamma_{S,\alpha}(g^2)\,,
\end{align}
which receives an anomalous contribution $\gamma_{S,\alpha}(g^2)$. We use index
$\alpha$ here to indicate that there exist few operators carrying the same Lorentz spin
$S$.

In virtue of conformal symmetry, the operators $ {\mathcal O}_\alpha(0)$ can be classified
according to representation of the $SO(2,4)$ conformal group (for a review, see  e.g.
\cite{Braun:2003rp}). For the light-ray operators \re{O-def}, the conformal symmetry
reduces to its collinear $SL(2)$ subgroup. This subgroup leaves the light-ray $x^\mu=z
n^\mu$ invariant and acts on the light-cone coordinates $z$ as
\begin{align}\label{Moeb}
z \to {a z+b \over c z +d} \,,\qquad ad-bc=1\,.
\end{align}
The corresponding transformation properties of the operator $ {\mathcal O}_\alpha\equiv
\mathcal{O}_{S,\alpha}$ are
\begin{align}\label{sl2}
\mathcal{O}_{S,\alpha}(z n)\to (cz+d)^{-2j_{S,\alpha}} \mathcal{O}_{S,\alpha}\lr{{a z+b \over c z +d}n}\,,
\end{align}
where the conformal spin $j_{S,\alpha}$ is related to the Lorentz spin of the operator and
its scaling dimension as
\begin{align}
j_{S,\alpha} = \frac12 \lr{S+ \Delta_{S,\alpha}} = S + \frac12 L + \frac12 \gamma_{S,\alpha}(g^2)\,.
\end{align}
The generators of the $SL(2)$ transformations \re{sl2} take the form of linear
differential operators acting on the light-cone coordinates of the operators
\begin{align}\notag\label{sl2-rep}
& L_-\,  \mathcal{O}_{S,\alpha}(z n) = -\partial_z\, \mathcal{O}_{S,\alpha}(z n)\,,
\\[1.5mm]\notag
& L_0\,  \mathcal{O}_{S,\alpha}(z n) = ( z\partial_z + j_{S,\alpha}(g^2))\mathcal{O}_{S,\alpha}(z n)\,,
\\[1.5mm]
& L_+\,  \mathcal{O}_{S,\alpha}(z n) = ( z^2\partial_z + 2 z j_{S,\alpha}(g^2))\mathcal{O}_{S,\alpha}(z n)\,.
\end{align}
It is straightforward to verify that the $SL(2)$ generators defined in this way satisfy
the standard commutation relations $[L_0,L_\pm] = \pm L_\pm$ and $[L_+,L_-] = 2L_0$. The
dependence of the last two relations in \re{sl2-rep} on the coupling constant reflects the
fact that the conformal generators $L_0$ and $L_+$ are modified by perturbative
corrections. The generator $L_-$ is related to the light-cone component of the total momentum
operator and is protected from loop corrections.

The operator $\mathcal{O}_{S,\alpha}(zn)$ belongs to the $SL(2)$ representation labelled
by the conformal spin $j_{S,\alpha}$.
As follows from \re{sl2-rep}, the operator $\mathcal{O}_{S,\alpha}(0)$ defines the lowest
weight of this representation and its descendants are given by total derivatives
 $(L_-)^\ell \mathcal{O}_{S,\alpha}(0)=(-\partial_z)^\ell  \mathcal{O}_{S,\alpha}(z n)\big|_{z=0}$. The conformal symmetry allows us to
 organise the sum on the right-hand side of \re{imp} as
 the sum over different $SL(2)$ moduli
\begin{align}\label{imp1}
\mathbb{O}({\bm z}) = \sum_{S,\alpha} \big[ \Phi_{S,\alpha}({\bm z})   {\mathcal O}_{S,\alpha}(0)+\text{descendants} \big]\,,
\end{align}
where `descendants' denote the contribution of the operators $(L_-)^\ell
\mathcal{O}_{S,\alpha}(0)$ and  the index $\alpha$ enumerates the conformal primary
operators ${\mathcal O}_{S,\alpha}$ with the same Lorentz spin $S$. The conformal symmetry
also fixes (up to an overall normalization) the two-point correlation function of these
operators
\begin{align}\label{2pt}
\vev{\mathcal{O}_{S,\alpha}(x) \mathcal{O}_{S',\alpha'}(0)} \sim \delta_{SS'} \delta_{\alpha\alpha'}\frac{(xn)^{2S}}{(x^2)^{S+\Delta_{S,\alpha}}} \,,
\end{align}
with the scaling dimension $\Delta_{S,\alpha}$ given by \re{Delta}.

Let us now consider the correlation function $\vev{\mathbb{O}({\bm z})
\mathcal{O}_{S,\alpha}(x)}$. Replacing the light-ray operator with its expansion \re{imp1}
and making use of \re{2pt}, we find that the correlation function receives a nonzero contribution from the operators 
$(L_-)^\ell \mathcal{O}_{S,\alpha}(0)$ (with $\ell=0,1,\ldots$) in \re{imp1} belonging
to the same $SL(2)$ moduli as $\mathcal{O}_{S,\alpha}(x)$. Moreover, taking the limit $x\to\infty$ we find that the leading
contribution only comes from the conformal primary operator $(\ell=0)$ whereas the
contribution of descendants is suppressed by factor of $1/|x|^\ell$ leading to
\begin{align}\label{int1}
\vev{\mathbb{O}({\bm z})  \mathcal{O}_{S,\alpha}(x)} \stackrel{x\to\infty}{\sim} \Phi_{S,\alpha}({\bm z}) \frac{(xn)^{2S}}{(x^2)^{S+\Delta_{S,\alpha}}} \,.
\end{align}
Thus
 the polynomial $\Phi_{S,\alpha}({\bm z})$ defines the leading asymptotic behavior
of the correlation function  $\vev{\mathbb{O}({\bm z})  \mathcal{O}_{S,\alpha}(x)}$ at
large distance  $x\to\infty$.

\subsection{Form factors}


Let us examine matrix elements of the conformal primary operators
with respect to asymptotic (on-shell) states  $\vev{0|\mathcal{O}_{S,\alpha}(0)|P}$.\footnote{
More general matrix elements of the form $\vev{P_1|\mathcal{O}_{S,\alpha}(0)|P_2}$ can be obtained from $\vev{0|\mathcal{O}_{S,\alpha}(0)|P}$ by
allowing some particles inside  the state $\ket{P}$ to carry negative energy.}
Here the asymptotic state  $\ket{P}$ consists  of
a fixed number of massless particles (scalars, gauginos and gluons)
each carrying the on-shell momentum $P_i^\mu$, certain helicity charge and the color $SU(N)$ charge $T^{a_i}$. The total color charge of the state is zero, $\sum_i T^{a_i}=0$, and the total momentum equals
$P^\mu=\sum_i P_i^\mu$.

Since the operators $\mathcal{O}_{S,\alpha}(0)$ arise from the expansion of the light-ray operator \re{imp1} it is natural
 to introduce the following quantity
\begin{align}\label{F}
F({\bm z},P)=\vev{0| \mathbb{O}({\bm z})|P} = \sum_{S,\alpha}  \Phi_{S,\alpha}({\bm z}) \vev{0|\mathcal{O}_{S,\alpha}(0)|P}+O(P\cdot n)\,,
\end{align}
which
can be thought of as a generating function of the form factors $\vev{0|\mathcal{O}_{S,\alpha}(0)|P}$.
Making use of the orthogonality condition \re{orth} we find from \re{F}
\begin{align}\label{ff}
\vev{0|\mathcal{O}_{S,\alpha}(0)|P} & = \widebar\Psi_{S,\alpha} ({\bm \partial_z})F({\bm z},P)\big|_{{\bm z}=0}\,.
\end{align}
The last term on the right-hand side of \re{F} describes the contribution of the $SL(2)$ descendant operators.
Such operators involve total derivatives of $\mathcal{O}_{S,\alpha}(0)$ and their matrix elements
are proportional to the light-cone component of the total momentum,
$\vev{0| (L_-)^\ell \mathcal{O}_{S,\alpha}(0)|P}\sim (Pn)^\ell \vev{0|\mathcal{O}_{S,\alpha}(0)|P}$.
Therefore we can eliminate the contribution of conformal descendants to \re{F} by choosing $(Pn)=0$.
We shall make use of this fact later in the paper.

Let us consider \re{F} in the special case when the state $\ket{P}$ consists of $L$ scalars each carrying the light-like momentum
$P_i^\mu$ aligned along the same light-like direction $\bar n_\mu$ (with $\bar n^2=0$)
\begin{align}\label{config}
P_i^\mu = p_i \, \bar n_\mu\,,\qquad P^\mu = (p_1+\ldots+p_L) \bar n_\mu\,,
\end{align}
with $-\infty < p_i<\infty$. In what follows we shall denote such state as $\ket{\bm p}$.
Then, evaluating the matrix element $\vev{0| \mathbb{O}({\bm z})|P}$ to the lowest order in the coupling, 
we  can be replace scalar fields by plane waves $\vev{0|Z(n z)|P_i} = \e^{i (P_i n) z} T^{a_i}$ to get
\begin{align}\label{F-bare}
F \left({\bm z},\bm p\right)= \tr\lr{T^{a_1}\ldots T^{a_L}}\sum_{i=1}^L \e^{i p_1 z_i +\ldots i p_L z_{i+L-1}}
+\text{perm}\,,
\end{align}
where we put $(n\bar n)=1$ for simplicity. Here the sum ensures the symmetry of $F({\bm z},\bm p)$ under 
the cyclic shift of ${\bm z}$'s (with $z_{i+L}\equiv z_i$) and `perm' denote terms with permutations of momenta 
and colour indices of particles. They are needed to restore the Bose symmetry of $F \left({\bm z},\bm p\right)$.
  
Combining together \re{ff} and \re{F-bare}, we obtain the following expression for the form factor of the conformal primary operator
in the kinematical configuration \re{config}, to the leading order in the coupling
\begin{align}\notag\label{F-col}
 \vev{0|\mathcal{O}_{S,\alpha}(0)| {\bm p}} & =    \widebar\Psi_{S,\alpha} ({\bm \partial_z})F ({\bm z},\bm p)\big|_{{\bm z}=0}
 \\[2mm]
 & =   \left[i^S L \,  \widebar\Psi_{S,\alpha} ({\bm p}) \tr\lr{T^{a_1}\ldots T^{a_L}}
+\text{(perm)}\right]\,.
\end{align}
Here in the second relation we took into account that $\Psi_{S,\alpha}({\bm p})$ is a cyclically invariant homogenous polynomial in ${\bm p}=(p_1,\dots,p_L)$ of degree $S$. 
We conclude from \re{F-col} that the polynomial $\Psi_{S,\alpha}({\bm p})$ defines the form factor
$\vev{0|\mathcal{O}_{S,\alpha}(0)|P}$ in the multi-collinear kinematical configuration \re{config}.

\subsection{Duality}

The polynomials  $\Phi_{S,\alpha}({\bm z})$ and $\Psi_{S,\alpha}({\bm p})$ depend on two
different sets of variables: the former depends on the light-cone coordinates of scalar
fields, whereas the latter is a function of the conjugated light-cone momenta. They define
the wave function of the same $L-$particle state in the coordinate and momentum
representations, respectively, and satisfy the
orthogonality condition \re{orth}.%
\footnote{The orthogonality condition \re{orth} can be casted into the well-known quantum
mechanical form. To see this we notice that $\int dx\, \bar \phi_\alpha(x)
\phi_\beta(x)=\int dx\, \bar \phi_\alpha(x) \e^{x\partial_z}\phi_\beta(z)\big|_{z=0}=
\widebar\psi_\alpha(i\partial_z) \phi_\beta(z)\big|_{z=0}$, with $\psi_\alpha(k)$ being
the Fourier transform of $\phi_\alpha(x)$.}

We recall that the explicit form of the polynomial $\Psi_{S,\alpha}({\bm p})$,
Eq.~\re{Phi-def}, is determined by the eigenstates
of the mixing matrix in the $SL(2)$ sector. 
The polynomials $\Phi_{S,\alpha}({\bm z})$ can then be obtained from the orthogonality
condition \re{orth}. Since the  $\Psi_{S,\alpha}({\bm p})$ and $\Phi_{S,\alpha}({\bm z})$
are defined in the two different representations, 
we do not expect them to be related to each other in an obvious way. 
The main goal of the present paper is to show that, due to integrability  of the
dilatation operator in planar $\mathcal{N}=4$ SYM, there exists the following  
relation between the two polynomials to the leading order in the coupling
\begin{align}\label{main}
\Psi_{S,\alpha}({\bm p}) =\xi_{S,\alpha}\, \Phi_{S,\alpha}({\bm z}) \,,\qquad p_i = z_i-z_{i+1}\,,
\end{align}
where the proportionality factor
$\xi_{S,\alpha}$  depends on the quantum numbers of the state and periodicity condition 
$z_{i+L}=z_i$ is implied.   Extension of \re{main} beyond the leading order  will be
discussed in the forthcoming paper \cite{prep}.

Notice that the duality relation \re{main} is formulated for the conformal primary operators $\mathcal{O}_{S,\alpha}(0)$.
The reason for this is that \re{main} becomes trivial for the descendant operator $(\partial_+)^\ell \mathcal{O}_{S,\alpha}(0)$
(with $\ell \ge 1$) involving a power of the total light-cone derivative.  According to definition \re{O-hat},
the $\Psi-$poly\-no\-mial for such operator is given by $(\sum_ip_i)^\ell \Psi_{S,\alpha}({\bm p})$ and
 it vanishes upon substitution into the left-hand side of \re{main}. Then, the
duality relation implies that the corresponding $\xi-$factor on the right-hand side of \re{main} vanishes as well.

The duality relation (\ref{main}) leads to another interesting property of the polynomial
$\Phi_{S,\alpha}({\bm z})$. Let us consider the completeness condition \re{comp} and
substitute  $p_i=w_i-w_{i+1}$ (with $w_{L+1}=w_1$). Since $\sum_i p_i=0$, the conformal descendants produce a
vanishing contribution to the left-hand side of \re{comp}. Then, we apply the
duality relation \re{main} to get from \re{comp}
\begin{align}\label{mag1}
\sum_{S,\alpha}  \xi_{S,\alpha}\, \widebar\Phi_{S,\alpha}({\bm z})\Phi_{S,\alpha}({\bm w}) =  \frac1{L}\sum_{i=1}^L \exp\lr{w_{12} z_i + \ldots+ w_{L1} z_{i+L-1 }}\,,
\end{align}
where $w_{i,i+1}\equiv w_i-w_{i+1}$ and the expression on the right-hand side is invariant under translations and cyclic
shifts of ${\bm z}$ and ${\bm w}$. Since $\Phi_{S,\alpha}({\bm z})$ is a homogenous
polynomial of degree $S$, we can further simplify \re{mag1} as
\begin{align}\label{mag2}
 \sum_{\alpha}  \xi_{S,\alpha}\, \widebar\Phi_{S,\alpha}({\bm z})\Phi_{S,\alpha}({\bm w}) =  \frac1{L}\sum_{i=1}^L
 \frac1{S!}  \lr{w_{12} z_i + \ldots+ w_{L1} z_{i+L-1 }}^S\,.
 \end{align}
Here the sum on the right-hand side runs over the conformal primary operators carrying the same spin $S$.
  
We have demonstrated in the previous subsections that  $\Psi_{S,\alpha}({\bm p})$ and $\Phi_{S,\alpha}({\bm z})$
have a simple interpretation, 
Eqs.~\re{int1} and \re{F-col}, respectively.
Then, the duality relation \re{main} establishes the correspondence between
the correlation function $\vev{\mathbb{O}({\bm z})  \mathcal{O}_{S,\alpha}(x)}$ at large distances $x\to\infty$ 
and the form factor $ \vev{0|\mathcal{O}_{S,\alpha}(0)|{\bm p}}$ evaluated for the special configuration of the 
light-cone momentum $p_i=z_i-z_{i+1}$.

In the rest of the paper, we demonstrate the validity of the duality relation
\re{main} to the lowest order in the coupling and explain its relation to the integrability of dilatation operator. 

\section{Duality at the leading order}

The polynomials $\Psi_{S,\alpha}({\bm p})$ and $\Phi_{S,\alpha}({\bm z})$ admit a
perturbative expansion, e.g.
\begin{align}
 \Psi_{S,\alpha}({\bm p}) = \Psi^{(0)}_{S,\alpha}({\bm p}) + g^2\, \Psi^{(1)}_{S,\alpha}({\bm p})  + O(g^4)\,.
\end{align}
In what follows we shall restrict our consideration to the leading term
$\Psi^{(0)}_{S,\alpha}({\bm p})$. To simplify notations, we will not display the
superscript `$(0)$'. The corresponding operator \re{O-hat} is conformal primary at one
loop -- it does not mix with other conformal operators at one loop and diagonalizes
the   dilatation operator at order $O(g^2)$.

\subsection{Conformal Ward identity}

Let us start with reviewing the constraints imposed by the conformal symmetry on the
polynomials $\Psi_{S,\alpha}({\bm p})$ and $\Phi_{S,\alpha}({\bm z})$.

At the leading order in the coupling, the light-ray operator \re{O-def} is given by the
product of free scalar fields $Z(n z_i)$. Each of them transforms under the conformal
$SL(2)$ transformations according to \re{sl2} with the conformal weight  $j_Z=1/2$. Then,
we use the relation \re{int1} and require the correlation function $\vev{\mathbb{O}({\bm
z})  \mathcal{O}_{S,\alpha}(x)}$ to be invariant under the conformal transformations
generated by $L_-$ and $L_0$, Eq.~\re{sl2-rep}, to find that  the polynomial
$\Phi_{S,\alpha}({\bm z})$ has to satisfy the conformal Ward identity
\begin{align} \label{Ward1}
 \sum_i  \partial_{z_i}  \Phi_{S,\alpha}({\bm z}) =
\bigg(\sum_i z_i \partial_{z_i} -S \bigg)  \Phi_{S,\alpha}({\bm z}) =0 \,,
\end{align}
so that $\Phi_{S,\alpha}({\bm z})$ should be translationaly invariant homogenous
polynomial in ${\bm z}$'s of degree $S$. The Ward identity for the  $L_+$ generator leads
to the relation between $\Phi_{S,\alpha}({\bm z})$ and the polynomials corresponding to
the descendant operators.

To obtain analogous relations for  $\Psi_{S,\alpha}({\bm p})$, we examine transformation
properties of the both sides of \re{O-hat} under the $SL(2)$ transformations, $L_a \,
{\mathcal O}_{S,\alpha}(0) = \Psi_{S,\alpha} ({\bm \partial_z})L_a \,\mathbb{O}({\bm z})
\big|_{{\bm z}=0}$.  Here the $SL(2)$ generators $L_a$ act  additively on each scalar field inside
$\mathbb{O}({\bm z})$ and  are given by \re{sl2-rep} with  $j_{S,\alpha}$ replaced by the
conformal spin of a free scalar field $j_Z=1/2$. Then, we impose the conditions
$(L_0-j_{S,\alpha}) \,  {\mathcal O}_{S,\alpha}(0) = L_+ \,  {\mathcal O}_{S,\alpha}(0)
=0$ that follow from  \re{sl2-rep} to get
 \begin{align}  \label{Ward2}
\bigg(\sum_i p_i \partial_{p_i} -S \bigg)  \Psi_{S,\alpha}({\bm p}) =\sum_i \bigg(p_i \partial_{p_i}^2 + \partial_{p_i} \bigg) \Psi_{S,\alpha}({\bm p})= 0 \,.
\end{align}
We would like to emphasize that the relations \re{Ward1} and \re{Ward2} were obtained to
the lowest order in the coupling. To higher orders, only the second relation in \re{Ward2}
is modified by perturbative corrections.

We recall that the polynomials have to satisfy the orthogonality condition \re{orth},
\begin{align}\label{orth1}
\widebar\Psi_{S,\alpha} ({\bm \partial_z})\Phi_{S',\alpha'}({\bm z}) \big|_{{\bm z}=0} = \delta_{SS'}\delta_{\alpha\alpha'} \,.
\end{align}
As was already mentioned, this relation suggests that the two functions should be related
to each other by a Fourier like transformation. An unusual feature of this transformation
is that it  maps one polynomial satisfying \re{Ward1}
into another polynomial verifying \re{Ward2}. Its explicit form has been worked out in Ref.~\cite{Derkachov:1997qv} 
\footnote{The inverse relation takes the form of the Fourier transform $
  \Psi_{S,\alpha}({\bm p}) = \int [\mathcal{D} {\bm z}] \, \e^{ \widebar{\bm  z}\cdot  {{\bm p}}}
 { \Phi_{S,\alpha}( {\bm z})} \,,
$ where ${\bm z}\cdot  {{\bm p}}\equiv \sum_i z_i \,p_i$ and  integration goes over the
unit disk, $z_i\bar z_i\le 1$,  in the complex $z-$plane with the  $SU(1,1)$ invariant
measure of spin $j=1/2$.}
\begin{align}\label{Psi-P}
  {\Phi}_{S,\alpha}({\bm z}) =&\int_0^\infty \prod_{i=1}^L dp_i  \e^{-p_i}   \Psi_{S,\alpha}( p_1 z_1,\ldots,p_L z_L)
  =
   \Psi_{S,\alpha}(\partial_{\bm w}) \prod_{i=1}^L \lr{1- w_i z_i}^{-1}\bigg|_{{\bm w}=0}\,.
\end{align}
Replacing $\Psi_{S,\alpha}({\bm p})$  in \re{Psi-P} with its general expression
\re{Phi-def}, we obtain the following result for the polynomial ${\Phi}_{S,\alpha}({\bm
z})$
\begin{align}\label{Psi-g}
\Phi_{S,\alpha} ({\bm z}) = \sum_{{\bm k}} c_{{\bm k},\alpha} \, z_1^{k_1} \ldots z_L^{k_L}\,.
\end{align}
 We recall that  $\Phi_{S,\alpha}({\bm z}) $ should
be translationally invariant and, therefore, the coefficients $c_{{\bm k},\alpha}$ are not
independent. In addition, substituting \re{Phi-def} and \re{Psi-g} into  \re{orth1} we
find (for $S=S'$) that they have to satisfy the orthogonality condition
\begin{align}\label{c-norm}
\sum_{{\bm k} } \lr{c_{{\bm k},\alpha}}^* c_{{\bm k},\alpha'}= \delta_{\alpha\alpha'}\,,
\end{align}
where the sum runs over $L$ nonnegative integers $\bm k=(k_1,\dots,k_L)$ such that $\sum_i
k_i=S$.

As an example, let us consider length 2 operators. In this case, for $L=2$, the conformal
Ward identity \re{Ward1} fixes $\Phi_{S}({\bm z})$ up to an overall normalization
\begin{align}\label{L2}
\Phi_{S}({\bm z})=c \,(z_1-z_2)^S\,.
\end{align}
Comparison with \re{Psi-g} shows that the expansion coefficients are given by binomial
coefficients, $c_{{\bm k}} =c\, (-1)^k \lr{S \atop k}$. Their substitution into
\re{Phi-def} yields the polynomial $\Psi_S({\bm p})$ as
\cite{Makeenko:1980bh,Ohrndorf:1981qv}
\begin{align}\label{Phi-L=2}
 \Psi_S({\bm p}) = c\sum_{k=0}^S    \frac{(-1)^{S-k}  S! }{(k! (S-k)!)^2}p_1^k p_2^{S-k}
=\frac{c}{S!} (p_1+p_2)^S {\rm C}_S^{1/2}\lr{p_1-p_2\over p_1+p_2}\,,
\end{align}
where ${\rm C}_S^{1/2}(x)$ is the Gegenbauer polynomial and the normalization factor
$c=[(S!)^2/(2S)!]^{1/2}$ is fixed by
 \re{c-norm}. We can now test the
duality relation \re{main} for $L=2$. Replacing $p_1=-p_2=z_{12}$ on the right-hand side
of \re{Phi-L=2} we get
\begin{align}
 \Psi_S({\bm p})\big|_{p_i=z_{i,i+1}}  {=} c  \sum_{k=0}^S    \frac{   S! \, (z_1-z_2)^S}{(k! (S-k)!)^2} = {(2S)! \over (S!)^3} \Phi_{S}({\bm z})\,,
\end{align}
in a perfect agreement with \re{main}.

For $L\ge 3$ the conformal symmetry \re{Ward1} and \re{Ward2} is not sufficient to fix the
polynomials $\Phi_{S,\alpha}({\bm z})$ and $\Psi_{S,\alpha}({\bm p})$. To find them, we
have to use integrability of the dilatation operator in planar $\mathcal{N}=4$ SYM.

\subsection{Dilation operator at one loop}

The explicit form of the conformal operators ${\mathcal O}_{S,\alpha}(0)$ and their
anomalous dimensions $\gamma_{S,\alpha}$ can be obtained by diagonalizing the dilatation
operator in the $SL(2)$ sector of $\mathcal{N}=4$ SYM. Making use of \re{O-hat} and
\re{imp}, the corresponding spectral problem can be reduced to solving a Schrodinger like
equation for the polynomials $\Phi_{S,\alpha}({\bm z})$ and $\Psi_{S,\alpha}({\bm p})$,
e.g.
\begin{align}\label{Sch}
\mathbb{H}  \, {\Phi}_{S,\alpha}({\bm z}) = \gamma_{S,\alpha}(g^2)   {\Phi}_{S,\alpha}({\bm z})\,.
\end{align}
To the lowest order in the coupling, the dilatation operator $\mathbb{H}$ can be mapped
into a Hamiltonian of the $SL(2)$ Heisenberg spin $j=1/2$ chain of length $L$
\begin{align}\label{H-sum}
 \mathbb{H} =g^2\left[ H_{12} + \ldots+H_{L1}\right] + O(g^4)\,.
\end{align}
Here the two-particle kernels $H_{i,i+1}$ act locally on the
light-cone coordinates $z_i$ and $z_{i+1}$ of two neighboring particles and admit the
following representation  \cite{Balitsky:1987bk,Braun:1998id,Belitsky:1999qh,Braun:1999te}
\begin{align}\label{2H}
 H_{i,i+1}\phi(z_i,z_{i+1}) = \int_0^1\frac{d \tau}{\tau}\bigg[ 2 \phi(z_i,z_{i+1}) - \phi((1-\tau) z_i + \tau z_{i+1}, z_{i+1}) -
 \phi(z_i,(1-\tau) z_{i+1} + \tau z_{i}) \bigg],
\end{align}
with $\phi(z_i,z_{i+1})$ being a test function. This operator has a clear physical interpretation
-- it displays two particles with the coordinates $z_i$ and $z_{i+1}$ in the direction of
each other along the light-cone.

The Hamiltonian $\mathbb{H}$ defined in \re{H-sum} and \re{2H} maps a homogenous
polynomial in ${\bm z}$ of degree $S$ into another homogenous polynomial of the same
degree. Then, replacing ${\Phi}_{S,\alpha}({\bm z})$ in \re{Sch} with its general
expression  \re{Psi-g} and comparing the coefficients in front of different powers of
${\bm z}$'s on the both sides of  \re{Sch}, we can obtain a system of linear homogenous
equations for the expansion coefficients $c_{{\bm k},\alpha}$. The corresponding
characteristic equation yields a polynomial equation for $\gamma_{S,\alpha}$. Solving the
system we should also take into account that ${\Phi}_{S,\alpha}({\bm z})$ has to be both
cyclically and translationally invariant homogenous polynomial in ${\bm z}$ of degree $S$.
This leads to the additional selection rule for the possible solutions.

In this way, it is straightforward to solve \re{Sch} for lowest values of the total spin
$S$. For instance, for $S=0$, the Schrodinger equation \re{Sch} has a trivial solution
${\Phi}_{S=0}({\bm z}) =1$ and $ \gamma_{S=0}(g^2)=0$. The corresponding conformal
operator $\mathcal{O}_{S=0}=\tr[Z^L]$ is a half-BPS state and its anomalous dimension
vanishes to all loops. For $S=2$ and $L=2,3,4$ we find the following expressions
\begin{align}\notag\label{ex-Psi}
&  \Phi^{(L=2)}_2=c_2\,z_{12}^2\,,
\\\notag
&  \Phi^{(L=3)}_2= c_3\,\big[z_{12}^2+z_{23}^2+z_{31}^2\big]  \,,
\\
&  \Phi^{(L=4)}_{2,\pm}=c_{4,\pm} \big[z_{12}^2+z_{23}^2+z_{34}^2+z_{41}^2
+(1\pm \sqrt{5}) (z_{12}z_{34}+z_{23}z_{41})\big],
\end{align}
where $z_{ij}\equiv z_i-z_j$ and the normalisation factors are fixed by the condition
\re{c-norm} to be $c_2^2=1/6$, $c_3^2=1/24$ and $c_{4,\pm}^2=(3\mp \sqrt{5})/160$. The
corresponding one-loop anomalous dimensions are
\begin{align}\label{ad-ex}
\gamma^{(L=2)}_{2} = 6g^2\,,\qquad \gamma^{(L=3)}_{2} = 4g^2\,,\qquad \gamma^{(L=4)}_{2,\pm} =(5\pm\sqrt{5})g^2\,.
\end{align}
Finally, we apply the relations \re{Phi-def} and \re{Psi-g} to obtain the corresponding
expressions for the $\Psi-$polynomial
\begin{align}\notag\label{ex-Phi}
&   \Psi^{(L=2)}_2 =c_2\big(\ft12 p_1^2 -2 p_1 p_2 +\ft12 p_2^2 \big),
\\[2mm] \notag
&   \Psi^{(L=3)}_2= c_3\big[   p_1^2+p_2^2+p_3^2 -2(p_1p_2+p_2p_3+p_3p_1) \big],
\\[2mm]\notag
&  \Psi^{(L=4)}_{2,\pm}=c_4\big[p_1^2+p_2^2+p_3^2+p_4^2  + 2(1\pm\sqrt{5}) (p_1p_3+p_2p_4)
\\
& \hspace*{45mm}- (3\pm \sqrt{5})(p_1p_2 +p_2p_3+p_3p_4+p_4p_1) \big].
\end{align}
By the construction, the polynomials \re{ex-Psi} and \re{ex-Phi} define the eigenstates of
the one-loop dilatation operator in $\mathcal{N}=4$ SYM in the two representations.

Let us verify that the obtained expressions satisfy the duality relation \re{main}. To
this end, we substitute $p_i=z_i-z_{i+1}$ (with $z_{i+L}\equiv z_i$) into the right-hand
side of \re{ex-Phi}  and compare the resulting expressions with  \re{ex-Psi}. We find
that, in a perfect agreement with \re{main}, the polynomials are indeed proportional to
each other,
\begin{align}
 \Psi_2^{(L)}(p_i=z_{i,i+1}) = \xi_{2}^{(L)} \Phi_2^{(L)}(z_i)\,,
\end{align}
 with the proportionality factor given (to the lowest order in the coupling) by
\begin{align}\label{xi-ex}
 \xi_{2}^{(L=2)} =  3\,,\qquad \xi_{2}^{(L=3)} =2 \,,\qquad \xi_{2,\pm}^{(L=4)} =  \frac{5\pm \sqrt{5}}{2}\,.
\end{align}

For higher values of $S$ and $L$, it is more efficient to construct the solution to
\re{Sch} with a help of Algebraic Bethe Ansatz \cite{FST,KS,Skl,Fad}. In this method,
the $\Phi-$ and $\Psi-$polynomials coincide with the  Bethe states in the $z-$ and $p-$representations,
respectively. Examining their explicit expressions  one can
verify that the duality relation \re{main} holds for arbitrary $S$ and $L$.

\section{Dual symmetry from the Baxter $Q-$operator}

We have demonstrated in the previous section that the one-loop dilatation operator
possesses the dual symmetry \re{main}. To understand the origin of this symmetry we use
the above mentioned relation between the one-loop dilatation operator and the
$SL(2;\mathbb{R})$ Heisenberg spin chain. The latter model is exactly solvable and its
eigenspectrum can be obtained using different technique \cite{FST,KS,Skl,Fad}. For our
purposes it is convenient to employ the method based on the Baxter $Q-$operator.

\subsection{Baxter $Q-$operator}

The method relies on the existence of the operator $\mathbb{Q}(u)$ which encodes an
information about all conserved charges of the $SL(2;\mathbb{R})$ Heisenberg spin chain.
It acts on the space of polynomials ${\Phi}({\bm z})$, depends on an arbitrary complex
parameter $u$ and satisfies the following defining relations
\begin{align}\notag\label{Q-def}
& [  \mathbb{Q}(u), \mathbb{Q}(v) ] =[  \mathbb{Q}(u), T(v) ] = 0 \,,
\\[2mm]
&  \mathbb{Q}(u+i) ( u+i/2)^L + \mathbb{Q}(u-i) ( u-i/2)^L=T(u) \, \mathbb{Q}(u)  \,,
\end{align}
where $T(u)=2u^L+ q_2 u^{L-2} + \ldots+q_L$ is the so-called auxiliary transfer matrix and
$q_k$ are commuting conserved charges. Then, to the lowest order in the coupling, the
Schr\"odinger equation \re{Sch} can be replaced with an analogous equation for
the Baxter ${Q}-$operator
\begin{align}\label{B-pr}
\mathbb{Q}(u) \, {\Phi}_{S,\alpha}({\bm z})  = Q_{S,\alpha}(u) \, {\Phi}_{S,\alpha}({\bm z}) \,.
\end{align}
The solutions to \re{B-pr} are characterized by the complete set of the conserved charges
$q_2,\ldots,q_L$. This allows us to identify index $\alpha$ with the set of their
eigenvalues  $\alpha=(q_2,\ldots,q_L)$.

For the $SL(2;\mathbb{R})$ magnet of an arbitrary length $L$ and (positive half-integer)
spin $j$,  the Baxter operator has been constructed in Ref.~\cite{Derkachov:1999pz}. For
our purposes we need its expression for $j=1/2$ spin
\begin{align}\label{bax}\notag
& \mathbb{Q}(u) \Phi(z_1,\ldots,z_L) =  c _Q \big[ \Gamma(iu+\ft12)\Gamma(-iu+\ft12)\big]^{-L}
\\
&\quad \times
\int_0^1\prod_{i=1}^L d\tau_i\,
\tau_i^{-iu-1/2} (1-\tau_i)^{iu-1/2}\, \Phi(\tau_1 z_1 + (1-\tau_1) z_{2},\ldots, \tau_L z_L + (1-\tau_L) z_{1})\,,
\end{align}
where $c_Q$ is a normalization factor and $\Phi(z_1,\ldots,z_L)$ is a test function. This
operator satisfies the following relations
\begin{align}\label{quasi}
\mathbb{Q}(i/2)= c_Q\,,\qquad \mathbb{Q}(-i/2)=c_Q\, \mathbb{P}\,,
\end{align}
where $\mathbb{P}$ is the operator of cyclic shift,  $z_i\mapsto z_{i+1}$. The Hamiltonian
$\mathbb{H}$, or equivalently the one-loop dilatation operator \re{H-sum},  is related to
the expansion of $\mathbb{Q}(u)$ around $u=\pm i/2$
\begin{align}\label{log-Q}
\mathbb{H} =g^2 \bigg[ i \big(\ln \mathbb{Q}(i/2)\big)' - i \big(\ln \mathbb{Q}(-i/2)\big)' \bigg] + O(g^4)\,,
\end{align}
where prime denotes a derivative with respect to the spectral parameter $u$.

The Schr\"odinger equation \re{Sch} is equivalent to the spectral problem \re{B-pr} for
the operator \re{bax}. As follows from the second relation in \re{Q-def}, the eigenvalues
of the operator $Q_{S,\alpha}(u)$ satisfy the second-order finite-difference equation, the
so-called $TQ-$relation. Having solved this equation, we can evaluate the one-loop
anomalous dimension from \re{log-Q} as
\begin{align}\label{logQ}
\gamma_{S,\alpha} =  g^2\bigg[ i \lr{\ln  {Q_{S,\alpha}}(i/2)}' - i \lr{\ln  {Q_{S,\alpha}}(-i/2)}' \bigg] + O(g^4)\,.
\end{align}
In addition, the condition for ${\Phi}_{S,\alpha}({\bm z})$ to be cyclic invariant
function of $z_i$ leads to the selection rule for $Q_{S,\alpha}(u)$
\begin{align}\label{cyc}
{Q_{S,\alpha}(-i/2) =  Q_{S,\alpha}(i/2)}\,,
\end{align}
which follows from $(\mathbb{P}-1){\Phi}_{S,\alpha}({\bm z}) =0$ combined with \re{quasi}.

It is straightforward to verify that the eigenstates ${\Phi}_{S,\alpha}({\bm z})$ found in
the previous section, indeed diagonalize the Baxter $Q-$operator \re{bax}. For $S=0$, we
substitute ${\Phi}_{S=0}({\bm z})=1$ into \re{B-pr} and \re{bax} to find $Q_{S=0}(u)=c_Q$.
For $S=2$, we use the relation \re{ex-Psi} to get the corresponding eigenvalues
\begin{align}\notag\label{Q-ex}
Q_{2}^{(L=2)}(u) &= -{1\over 12} + u^2\,, && \left(q_2=-\ft{13}2 \right),
\\[2mm] \notag
Q_{2}^{(L=3)}(u) &= -{1\over 4} + u^2\,, && \left(q_2=-\ft{19}2,\ q_3=0\right),
\\
Q_{2,\pm}^{(L=4)}(u) &= -{1\over 4} \pm {\sqrt{5} \over 10} + u^2\,, && \left(q_2=-13,\ q_3= 0,\ q_4 = \ft{21} 8\pm \sqrt{5}\right).
\end{align}
Here we also indicated the corresponding values of the conserved charges $q_2,\ldots,q_L$.
They are uniquely fixed
by the $TQ-$relation \re{Q-def}.%
\footnote{It follows from the $TQ-$relation at large $u$ that $q_2=-(S+L/2)(S+L/2-1) -
L/4$. } Also, the expressions \re{Q-ex} satisfy the normalization condition $Q(u)\sim u^S$
at large $u$ which fixes the constant $c_Q$ in \re{bax}. The substitution of \re{Q-ex}
into \re{logQ} yields the correct result for the one-loop anomalous dimensions \re{ad-ex}.

For  arbitrary total spin $S$ and length $L$, the eigenvalues of the $\mathbb{Q}-$operator \re{bax}
are given by polynomials in $u$ of degree $S$
\begin{align}\label{uk}
Q_{S,\alpha}(u) = \sum_{k=0}^S c_{k} u^k = \prod_{k=1}^S (u - u_{k})\,,
\end{align}
with the expansion coefficients $c_{k}$ and roots $u_k$ depending on the conserved
charges. The relation \re{uk} corresponds to the particular choice of the normalization
factor $c_Q$ in \re{bax}. We find from the first relation in \re{quasi} that it is given
by
\begin{align}\label{cQ}
c_Q= Q_{S,\alpha}(i/2) = \prod_{k=1}^S \left(\ft{i}2 - u_{k}\right).
\end{align}
Requiring  $Q_{S,\alpha}(u)$ to satisfy the $TQ-$relation \re{Q-def}, we find (by putting
$u=u_k$ on both sides of \re{Q-def}) that the parameters $u_k$ have to satisfy the Bethe
root equations for the $SL(2)$ Heisenberg magnet of spin $1/2$
\begin{align}
\lr{u_k+i/2 \over u_k-i/2}^L = \prod_{n\neq k} {u_k-u_n-i \over u_k-u_n+i}\,.
\end{align}
For the $SL(2;\mathbb{R})$ magnet the roots $u_k$ take real values only.

\subsection{Large $u$ expansion}

In this subsection, we show following Ref.~\cite{Derkachov:1999pz} that the dual
symmetry can be derived from the asymptotic expansion of both sides of \re{B-pr} at large
$u$. In this limit, the eigenvalues of the Baxter $Q-$operator \re{uk} scale as
\begin{align}\label{large}
 Q_{S,\alpha}(u) = u^S \left[ 1 - {\sum_k u_k \over u} + O(1/u^2)\right].
\end{align}
This suggests that the operator $\mathbb{Q}(u)$ should admit similar expansion
\begin{align}\label{Q0}
 \mathbb{Q}(u) {\Phi}_{S,\alpha}({\bm z}) = u^S \left[ \mathbb{Q}^{(0)} + \frac1{u} \,\mathbb{Q}^{(1)} + O(1/u^2)\right]  {\Phi}_{S,\alpha}({\bm z})\,,
\end{align}
with $\mathbb{Q}^{(0)}, \mathbb{Q}^{(1)}, \ldots$ being mutually commuting operators.
Substituting \re{large} and \re{Q0} into \re{B-pr}, we compare the coefficients in front
of powers of $u$ to find
\begin{align}\notag\label{rel1}
 \mathbb{Q}^{(0)} {\Phi}_{S,\alpha}({\bm z}) &=  {\Phi}_{S,\alpha}({\bm z})\,,\qquad
 \\[2mm]
 \mathbb{Q}^{(1)} {\Phi}_{S,\alpha}({\bm z}) &=  - \big(\sum_k u_k\big) {\Phi}_{S,\alpha}({\bm z})\,,\quad \ldots
\end{align}

To get explicit expression for the operators $\mathbb{Q}^{(0)}, \mathbb{Q}^{(1)}, \ldots$,
we apply \re{Psi-P} and replace a test polynomial $\Phi({\bm z})$ on the right-hand side
of \re{bax} with its expression in terms of  polynomial $\Psi({\bm p})$,
\begin{align}\label{lift}
\Phi({\bm z}) =  \Psi(\partial_{\bm w}) \prod_{i=1}^L \lr{1- w_i z_i}^{-1}\bigg|_{{\bm w}=0}\,.
\end{align}
Then, integration over $\alpha-$parameters in \re{bax} yields another (equivalent)
representation for the $Q-$operator
\begin{align}
 \mathbb{Q}(u) \Phi({\bm z}) =c_Q \,\Psi(\partial_{\bm w})\prod_{i=1}^L (1-w_iz_i)^{iu-1/2} (1-w_iz_{i+1})^{-iu-1/2}\bigg|_{{\bm w}=0}\,.
\end{align}
Finally, we rescale the auxiliary parameters $w_i\to w_i/u$ and expand the product on the
right-hand side in powers of $1/u$ to get after some algebra
\begin{align}\label{Q-test}
   \mathbb{Q}(u) \Phi({\bm z}) =  c_Q \bigg[1 +\frac{i}{2u} \sum_{i=1}^L (z_i+z_{i+1}) \lr{p_i\partial_{p_i}^2+\partial_{p_i} }
   + O(u^{-2})\bigg]\Psi(-iu{\bm p})
  \bigg|_{p_i=z_{i,i+1}}\,,
\end{align}
where we used a shorthand notation for $c\,{\bm p}\equiv (cp_1,\ldots,cp_L)$.  
To match \re{Q0}, we replace test functions in \re{Q-test}  by the eigenfunctions $\Phi_{S,\alpha}$ and $\Psi_{S,\alpha}$
to get 
\begin{align}\notag\label{rel2}
 \mathbb{Q}^{(0)} {\Phi}_{S,\alpha}({\bm z}) &= (-i)^S c_Q \,\Psi_{S,\alpha}({\bm p}) \big|_{p_i=z_{i,i+1}}\,,
 \\[2mm]
 \mathbb{Q}^{(1)} {\Phi}_{S,\alpha}({\bm z}) &=\frac12(-i)^{S-1} c_Q \sum_{i} (z_i+z_{i+1}) \lr{p_i\partial_{p_i}^2+\partial_{p_i} }\,\Psi_{S,\alpha}({\bm p}) \big|_{p_i=z_{i,i+1}}\,,\quad \ldots
\end{align}
As follows from the first relation, the
 operator $\mathbb{Q}^{(0)}$ generates the duality
transformation -- it transforms the polynomial ${\Phi}_{S,\alpha}({\bm z}) $ from the
coordinate to momentum representation and assigns the light-cone momenta as
$p_i=z_i-z_{i+1}$. Since ${\Phi}_{S,\alpha}({\bm z}) $ diagonalizes the operator
$\mathbb{Q}^{(0)}$, Eq.~\re{rel1}, we conclude from \re{rel2} that
it has to satisfy the duality relation \re{main} with
\begin{align}
\xi_{S,\alpha} =  i^S / c_Q = i^S/Q_{S,\alpha}(i/2) = \bigg[\prod_{k=1}^S (\ft12 +i u_{k})\bigg]^{-1}\,,
\end{align}
where  the second relation follows from \re{cQ}. We verified that for $S=2$ this relation is
in an agreement with  \re{xi-ex} and \re{Q-ex}.


In the similar manner, we can equate the expressions for $\mathbb{Q}^{(1)}
{\Phi}_{S,\alpha}({\bm z})$ (as well as for the remaining subleading terms in the large
$u$ expansion of the Baxter operator) on both sides of \re{rel1} and \re{rel2}
to obtain the additional relations between $\Psi_{S,\alpha}({\bm p})$ and
${\Phi}_{S,\alpha}({\bm z})$. As we show in the next subsection, for the polynomials verifying 
\re{main}, these relations are automatically fulfilled.

\subsection{Dual symmetry at work}

Let us understand to which extend the dual symmetry \re{main} fixes the polynomials
$\Psi_{S,\alpha}({\bm p})$ and $\Phi_{S,\alpha}({\bm z})$. Since the relation \re{main}
involves $\Psi_{S,\alpha}({\bm p})$ evaluated for $p_i=z_i-z_{i+1}$, we might expect that
it will allow us to obtain $\Psi_{S,\alpha}({\bm p})$ for the  vanishing total momentum
only,  $\sum_i p_i=0$. As we will see in a moment, this is not the case -- the dual
symmetry determines the polynomial $\Psi_{S,\alpha}({\bm p})$ for arbitrary total
momentum.

To begin with, we apply \re{Psi-P} and rewrite the duality relation \re{main} as
\begin{align}\label{dual-rel}
\Psi_{S,\alpha}(p_1,\ldots,p_L)\bigg|_{p_i=z_i-z_{i+1}} =\xi_{S,\alpha}\, \int_0^\infty \prod_{i=1}^L {
dt_i \e^{-t_i}}\, \Psi_{S,\alpha}(t_1z_1,\ldots,t_Lz_L) \,.
\end{align}
In general, $\Psi_{S,\alpha} ({\bm p})$ is a homogenous polynomial in $ {\bm
p}=(p_1,\dots,p_L)$ of degree $S$, Eq.~\re{Phi-def}.
It is uniquely specified by the set of the expansion coefficients $c_{{\bm k},\alpha}$
invariant under the cyclic shift of integers ${\bm k}=(k_1,\ldots,k_L)$. Substituting
\re{Phi-def} into both sides of \re{dual-rel} we arrive at
\begin{align}\label{c-sys}
\sum c_{{\bm k},\alpha}\,{ (z_1-z_2)^{k_1}\over k_1!} \ldots {(z_{L}-z_1)^{k_L}\over k_L!} = \xi_{S,\alpha}\, \sum c_{{\bm k},\alpha}\, z_1^{k_1}\ldots z_L^{k_L}\,,
\end{align}
where the sum on both sides runs over nonnegative integers satisfying $k_1+\ldots+k_L=S$.

Analysis of the equation \re{c-sys} goes along the same lines as in the beginning of
Sect.~3.2. Namely, we compare coefficients in front of powers of ${\bm z}$'s on both sides
of \re{c-sys} and obtain the system of linear homogenous equations for the coefficients
$c_{{\bm k},\alpha}$. The characteristic equation for this system yields a polynomial
equation for $\xi_{S,\alpha}$. Then, each solution $\xi_{S,\alpha}$ leads to a definite
expression for $c_{{\bm k},\alpha}$ (modulo an overall normalization)
which we can use to determine the polynomial 
$\Psi_{S,\alpha} ({\bm p})$, Eq.~\re{Phi-def}, for arbitrary total momentum $\sum_i p_i$.
The overall normalization of $c_{{\bm k},\alpha}$ is fixed by \re{c-norm}. It is
straightforward to verify that for $S=2$ and $L=2,3,4$ the solutions to \re{c-sys}
obtained in this way coincide with those in \re{ex-Phi}.

Thus the duality relation \re{dual-rel}  is powerful enough to determine the eigenstates
of the one-loop dilatation operator \re{Sch}. According to \re{B-pr}, the same polynomials
diagonalize the Baxter operator $\mathbb{Q}(u)$ for arbitrary $u$. Then,  it follows from
\re{Q0} that at large $u$ solutions to \re{dual-rel} diagonalize the operators
$\mathbb{Q}^{(0)}, \mathbb{Q}^{(1)},\ldots$ defined in \re{rel2}.

\section{Supersymmetric extention of duality}

In this section, we extend the duality symmetry to supersymmetric light-ray operators
built from various components of gaugino and gauge strength fields in $\mathcal{N}=4$ SYM.

\subsection{Light-cone superfields and superstates}

Supersymmetric  light-ray operators have the same form as \re{O-def} with 
complex scalar field $Z(x)$ replaced by the (chiral) super field (see Eq.~\re{super-O} below)
\begin{align}\label{superZ}
\mathcal{Z}(x,\theta)=\e^{\theta_a Q_n^a} Z(x) = Z(x) + \theta_a   \psi^a_n(x)
+ \frac12 \theta_a\theta_b \epsilon^{ab} F_{n}(x)\,.
\end{align}
Here $\theta^a$ are odd (Grassmann) coordinates and  the generators of supersymmetric
transformations $Q_n^a$ (with $a=1,2$) are given by two linear combinations  of
$\mathcal{N}=4$ supercharges
\begin{align}\label{Q-proj}
Q_n^a = \lambda^\alpha (n) \,Q_\alpha^a\,,\qquad n_{\alpha\dot\alpha} = \lambda_\alpha(n) \tilde \lambda_{\dot\alpha}(n)\,,
\end{align}
with $n_{\alpha\dot\alpha} = n_\mu (\sigma^\mu)_{\alpha\dot\alpha}$ (with
$\alpha,\dot\alpha=1,2$) being the light-like vector entering the definition of the
light-ray operator \re{O-hat}.   The operators $\psi^a_n(x)$ and $F_{n}(x)$ are given by
linear combinations of gaugino field, $\psi_\alpha^a(x)$, and self-dual part of the
strength tensor, $F_{\alpha\beta}(x)$,
\begin{align}
 \psi^a_n(x) =\lambda^\alpha(n) \,\psi_\alpha^a(x)\,,\qquad F_{n}(x)=\lambda^\alpha(n)\lambda^\beta(n)F_{\alpha\beta}(x)\,.
\end{align}
Notice that the right-hand side of \re{superZ} does not involve all field components in
$\mathcal{N}=4$ SYM. The reason for this is that constructing the superfield \re{superZ}
we only used two (out of four) supercharges. In fact, as we show in Appendix~A, the
operator $\mathcal{Z}(x,\theta)$ defines the $\mathcal{N}=2$ part of the  full
$\mathcal{N}=4$ light-cone superfield. The latter superfield contains spurious field
components (see Eq.~\re{Phi}) whose contribution should be carefully separated. To avoid
this complication we prefer to deal with the $\mathcal{N}=2$ superfield \re{superZ}.

To define form factors of supersymmetric light-cone operators, we also need a
supersymmetric extension of the on-shell asymptotic states. The operators $(Z(x),
\psi^a_n(x), F_{n}(x))$ create out of vacuum particles with helicity $h=(0, +1/2, +1)$,
respectively, e.g.
\begin{align} \label{sin}
  \vev{0 |\psi^a_\alpha(x) |p,\ft12,b } = \lambda_\alpha(p)\, \delta^{ab} \e^{ipx}\,,
\quad
  \vev{0 | F_{\alpha\beta}(x)  |p,1} = \lambda_\alpha(p)  \lambda_\beta(p) \e^{ipx}\,,
\end{align}
where $\lambda_\alpha(p)$ is defined by light-like momentum of particles
$p_{\alpha\dot\alpha} = \lambda_\alpha(p) \tilde\lambda_{\dot\alpha}(p)$. To simplify
formulae we do not display here color indices. In a close analogy with \re{superZ}, we can
combine the single-particle states $\ket{p,h}$  into a single $\mathcal{N}=2$ superstate
by introducing odd variables $\eta^a$ (with $a=1,2$)
\begin{align}\notag\label{super-state}
\ket{p,\eta} &= \big[a_{0}^\dagger(p)+\eta^a a_{1/2,a}^\dagger(p) + \ft12 \epsilon_{ab}\eta^a\eta^b \, a_{1}^\dagger(p)  \big] \ket{0}
\\[2mm]
&
\equiv \ket{p,0} +\eta^a \ket{p,\ft12,a} +\ft12 \epsilon_{ab}\eta^a\eta^b  \ket{p,1}\,.
\end{align}
Here $a^\dagger_h(p)$ is the creation operator of a massless particle with light-like
momentum $p^\mu$ and helicity $h$. To equate the helicity charge of three terms on the
right-hand side of \re{super-state}, we assign helicity $(-1/2)$ to odd variables
$\eta^a$. As before, the superstate $\ket{p,\eta}$  describes $\mathcal{N}=2$ part of the
full  $\mathcal{N}=4$ supermultiplet of on-shell states.

We combine together relations \re{superZ}, \re{sin}  and  \re{super-state}  to find
\begin{align}\label{super-wave}
\vev{0|\mathcal{Z}(x,\theta)|p,\eta } =\big[1+\vev{np} (\theta\eta) + \ft12 \vev{np}^2 (\theta\eta)^2 \big] \e^{ipx}
=\e^{ipx +\vev{np}(\theta\eta)  }\,,
\end{align}
where $\vev{np} =\lambda^\alpha(n) \lambda_\alpha(p)$ and $(\theta\eta)=\theta_a \eta^a$.
To define the supermomentum carried by the superfield, we perform Fourier transformation
of \re{super-wave} with respect to $x^\mu$ and $\theta^a$
\begin{align}
\int d^4 x\, \e^{-ikx}\int d^2 \theta \, \e^{-(\theta\vartheta)}\vev{0|\mathcal{Z}(x,\theta)|p,\eta } = (2\pi)^4 \delta^{(4)}(k-p)
\delta^{(2)}(\vartheta- \vev{np} \eta)\,.
\end{align}
We conclude from this relation that the odd momenta carried by the superfield and the
superstate \re{super-state} are related to each other as $\vartheta^a=\vev{np} \eta^a$.

\subsection{Light-ray operators in superspace}

Supersymmetric generalization of the light-ray operator \re{O-hat} looks as
\begin{align}\label{super-O}
\mathbb{O}({\bm Z}) = \tr\left[ \mathcal{Z}(z_1 n,\theta_1) \ldots Z(z_L n,\theta_L)\right]\,,
\end{align}
where  ${\bm Z}=(z_1,\theta_1),\ldots,(z_L,\theta_L)$ are the coordinates of $L$
supefields in the light-cone superspace. As before, the gauge invariance can be restored
by inserting the light-like Wilson lines between the adjacent super fields. Similar to
\re{start}, the expansion of $\mathbb{O}({\bm Z})$ in powers of $z$'s produces local
operators of length $L$ containing an arbitrary number of covariant derivatives. Further
expansion in powers of $\theta$'s yields operators $\tr[D_+^{k_1}X_1\ldots D_+^{k_L} X_L]$
of different partonic content $X=(Z,\psi^a_n,F_n)$.

Different components of the superfield \re{superZ} are transformed under the conformal
$SL(2)$ transformations according to \re{sl2} and carry the conformal spin, $j_Z=1/2$,
$j_\psi=1$ and $j_F=3/2$. When combined together, they define the so-called atypical, or
chiral, representation of the superconformal $SL(2|2)$ group \cite{Belitsky:2004sc}.
Namely, the light-cone superfield $\mathcal{Z}(z n,\theta)$   transforms  linearly under
the $SL(2|2)$ transformations
\begin{equation}\label{local}
\delta_G \mathcal{Z}(z n,\theta) = G \, \mathcal{Z}(z n,\theta)\,,
\end{equation}
with the generators $G=\{L^\pm, L^0, {W}{}^{a,\pm}, {V}^\pm_{a},{T}_b{}^a\}$ given by the
differential operators acting on its coordinates
\begin{align}\notag\label{sl2|2}
& {L}^- = -\partial_z \, , && {L}^+ =  z + z^2\partial_z + z \left( \theta\cdot
\partial_\theta \right) \, , && {L}^0 = \ft12 + z
\partial_z + \ft12\left( \theta\cdot \partial_\theta \right)\,,
\\[2mm]\notag
& {W}{}^{a,-} = \theta^a \, \partial_z \, , &&  {W}{}^{a,+} =
\theta^a [ 1 +  z \partial_z +  \left( \theta\cdot
\partial_\theta \right) ] \, ,  &&
{T}_b{}^a = \theta^a
\partial_{\theta^b} - \ft12 \, \delta_b^a \left( \theta\cdot
\partial_\theta \right) \, ,
\\[2mm]
& {V}^-_{a} = \partial_{\theta^a} \,, &&  {V}^+_{a} = z\partial_{\theta^a} \, ,&&
\end{align}
where $\partial_z \equiv \partial/\partial z$ and $\theta \cdot \partial_\theta \equiv
\sum_{a=1,2}\theta^a \partial/\partial\theta^a$. A global form of the transformations
\re{local} can be found in Ref.~\cite{Belitsky:2004sc}.

The super light-ray operator \re{super-O} belongs to the tensor product of $L$ copies of
the $SL(2|2)$ representations. Similarly to \re{imp1}, we can classify all possible local
operators entering the OPE expansion of $\mathbb{O}({\bm Z})$ according to irreducible
components that appear in this tensor product
\begin{align} \label{super-OPE}
\mathbb{O}({\bm Z}) = \sum_{S,\alpha} \big[  {\Phi}_{S,\alpha}({\bm Z})   {\mathcal O}_{S,\alpha}(0)+\text{descendants} \big]\,,
\end{align}
where the contribution of each component involves the superconformal primary operator
 ${\mathcal O}_{S,\alpha}(0)$ and its $SL(2|2)$ descendants. The polynomial $\Phi_{S,\alpha}({\bm Z})$ satisfies the lowest weight condition
\begin{align} \label{lw}
L^-\, \Phi_{S,\alpha}({\bm Z}) = W^{a,-}
\,\Phi_{S,\alpha}({\bm Z}) = V_a^-\, \Phi_{S,\alpha}({\bm Z}) = 0\,,
\end{align}
and diagonalizes the operators $L^0$ and $T_b{}^{a}T_a{}^{b}$. Here, the $SL(2|2)$
generators  are given by the sum of differential operators \re{sl2|2} acting on the
coordinates of $L$ particles. In general, $\Phi_S({\bm Z})$ is given by a homogenous
polynomial in ${\bm z}$'s and $\mathbf{\theta}$'s of the total degree $S$. For $L=2$ its
explicit expression can be found in Ref.~\cite{Belitsky:2005gr}. For $L\ge 3$ the
relations \re{lw} do not fix the polynomials $\Phi_{S,\alpha}({\bm Z})$ completely. The
additional condition for $\Phi_{S,\alpha}({\bm Z})$ comes from integrability of the
dilatation operator for the light-ray operators \re{super-O} which can be mapped, to the
lowest order in the coupling,  into Hamiltonian of the $SL(2|2)$  Heisenberg spin chain.

The operator  ${\mathcal O}_{S,\alpha}(0)$ can be obtained from \re{super-OPE} as
\begin{equation}\label{O4}
{\mathcal O}_{S,\alpha}(0) = \widebar\Psi_{S,\alpha}(\partial_{\bm Z})\, \mathbb{O}({\bm Z}) \big|_{{\bm Z}=0}\,,
\end{equation}
where $\partial_{\bm
Z}=(\partial_{z_1},\partial_{\theta_1}),\ldots,(\partial_{z_L},\partial_{\theta_L})$  and
$\Psi_{S,\alpha}({\bm P})$ is a polynomial in ${\bm
P}=(p_1,\vartheta_1),\ldots,(p_L,\vartheta_L)$. The two (super)polynomials,
$\Phi_{S,\alpha}({\bm Z})$ and $\Psi_{S,\alpha}({\bm P})$, uniquely determine the form of
the superconformal operators and their contribution to the OPE \re{super-OPE}. They
satisfy the relations
\begin{align}\notag
& \widebar\Psi_{S,\alpha}(\partial_{\bm Z}) \Phi_{S',\alpha'}({\bm Z})  \big|_{{\bm Z}=0} = \delta_{SS'}\delta_{\alpha\alpha'}\,,
\\
& \sum_{S,\alpha}\widebar\Psi_{S,\alpha}({\bm P}) \Phi_{S,\alpha}({\bm Z})  =\frac1{L}
\sum_{i=1}^L
\prod_{k=i}^{i+L-1}
 \exp\lr{ p_1z_k+\vartheta_1 \theta_k }
 \,,
\end{align}
(with $z_{k+L}=z_k$ and $\theta_{k+L}=\theta_k$) which generalise similar relations in the
$SL(2)$ sector, Eqs.~\re{orth} and \re{comp}. The solution to these relations reads
\cite{Belitsky:2005gr}
\begin{align}\notag\label{super-rel}
\Phi_{S,\alpha}(Z_1,\ldots,Z_L) & =  \int_0^\infty\prod_{k=1}^L dt_k\,  \e^{-t_k} \Psi_{S,\alpha}(t_1 Z_1,\ldots,t_L
Z_L)
\\
& = \Psi_{S,\alpha}(\partial_{W_1},\ldots,\partial_{W_L}) \prod_{k=1}^L (1-w_k  z_k
 -   \zeta_k\cdot\theta_k)^{-1}\big|_{w_k=\zeta_k=0}\,,
\end{align}
where $Z_i\equiv (z_i,\theta_i^a)$ and $\partial_{W_k} \equiv
(\partial_{w_k},\partial_{\zeta_k})$.

The relations \re{super-rel} are remarkably similar to those in the $SL(2)$ sector,
Eqs.~\re{Psi-P} and \re{lift}, and they are coincide for $\theta_i^a=0$. Indeed, the super
light-ray operator \re{super-O} reduces for $\theta_i^a=0$ to its lowest component given
by \re{O-hat}. This allows us to interpret the polynomial $\Phi_{S,\alpha}({\bm Z})$ as
being obtained from the analogous polynomial in the $SL(2)$ sector through the lift from
the light-cone to the superspace $z_i\mapsto (z_i,\theta_i)$ \cite{Belitsky:2004sc}.

\subsection{Baxter operator for the $SL(2|2)$ spin chain}

To lowest order in the coupling, the scale dependence of the super light-ray operators
\re{super-O} in planar $\mathcal{N}=4$ SYM can be determined from the eigenspectrum of the
$SL(2|2)$ Heisenberg spin chain. Namely, the polynomials $\Phi_{S,\alpha}({\bm Z})$
coincide with the eigenstates of the spin chain whereas the corresponding energies
determine the one-loop correction to the anomalous dimension of the superconformal
operators in \re{super-OPE}.

As in the case of the $SL(2)$ spin chain, we replace the Schr\"odinger equation for the
$SL(2|2)$ spin chain with the spectral problem for the Baxter $Q-$operator
\begin{align}\label{super-bax1}
\mathbb{Q}_i(u) \Phi_{S,\alpha}({\bm Z}) = Q_{S,\alpha,i}(u) \Phi_{S,\alpha}({\bm Z})\,.
\end{align}
Here we introduced the subscript $i=1,\ldots,4$ to indicate that there are few
$Q-$operators in this case.
The operators $\mathbb{Q}_i(u)$ act on the tensor product of $L$ copies of chiral
$SL(2|2)$ representation and satisfy the same defining relations as in the previous case.
Namely, the operators $\mathbb{Q}_i(u)$ commute with the $SL(2|2)$ transfer matrices and
among themselves  but the corresponding TQ relations are move involved in this case. A
general approach to constructing such operators has been developed in
Refs.~\cite{Belitsky:2006cp,Bazhanov:2008yc}.

For our purposes, we will only need an explicit expression for one of the $Q-$operators
which plays a special role in finding the eigenspectrum of the model -- it determines the
Hamiltonian of the $SL(2|2)$ magnet through the same relation as for the $SL(2)$ spin
chain, Eq.~\re{log-Q}. Denoting this operator by $\mathbb{Q}(u)$, we find that it acts on
a test polynomial $\Phi(Z_1,\ldots,Z_L)$ depending on $Z_i=(z_i,\theta_i^a)$ as
\begin{align}\label{super-bax}\notag
& \mathbb{Q}(u) \Phi(Z_1,\ldots,Z_L) =  c _Q \big[ \Gamma(iu+\ft12)\Gamma(-iu+\ft12)\big]^{-L}
\\
&\quad \times
\int_0^1\prod_{i=1}^L d\tau_i\,
\tau_i^{-iu-1/2} (1-\tau_i)^{iu-1/2}\, \Phi(\tau_1 Z_1 + (1-\tau_1) Z_{2},\ldots, \tau_L Z_L + (1-\tau_L) Z_{1})\,,
\end{align}
where $c_Q$ is a normalization factor and we used the shorthand notation for $\alpha Z_1 +
\beta Z_{2}=(\alpha z_1+\beta z_2,\alpha \theta_1+\beta \theta_2)$. This operator
satisfies the relations
\begin{align}
\mathbb{Q}(i/2)= c_Q\,,\qquad \mathbb{Q}(-i/2)=c_Q\, \mathbb{P}\,,
\end{align}
where $\mathbb{P}$ is the operator of cyclic shifts $(z_i,\theta_i)\mapsto
(z_{i+1},\theta_{i+1})$. Applying the second relation in \re{super-rel}, we can obtain
another, equivalent representation for the operator \re{super-bax}
\begin{align}\label{super-Q1}
\mathbb{Q}(u) \Phi({\bm Z}) =
c_Q \,\Psi(\partial_{\bm W})\prod_{i=1}^L (1-w_iz_i-\zeta_i\cdot \theta_i)^{iu-1/2} (1-w_iz_{i+1}-\zeta_i\cdot \theta_{i+1})^{-iu-1/2}\bigg|_{{\bm W}=0}\,,
\end{align}
where $\partial_{W_i} \equiv (\partial_{w_i},\partial_{\zeta_i})$. As before, we observe a
striking similarity of \re{super-bax} with the analogous expression \re{bax} for the
$SL(2)$ magnet. This is not accidental of course since for $\theta_i^a=0$ (with
$i=1,\ldots,L$) the eigenstate $\Phi_{S,\alpha}({\bm Z})$ reduces to its lowest component
which satisfies the $SL(2)$ Baxter equation \re{bax}.

Let us examine the asymptotic expansion of the operator $\mathbb{Q}(u)$ at large $u$. We
rescale $w_i\to w_i/u$ and $\zeta_i\to \zeta_i/u$ on the right-hand side of \re{super-Q1}
and go along the same lines as in \re{Q-test} to get
\begin{align} \label{super-Q2}
   \mathbb{Q}(u) \Phi({\bm Z}) =  c_Q \big[1 + O(u^{-1})\big]\Psi(-iu{\bm P})
  \big|_{P_i=Z_i-Z_{i+1}}\,,
\end{align}
where the $\Psi-$polynomial is evaluated for $P_i=(z_{i,i+1},\theta^a_{i,i+1})$. Finally,
we replace a test polynomial in \re{super-Q2} with the eigenstate of the $Q-$operator, use
the fact that $\Psi_{S,\alpha}({\bm P})$ is a homogenous polynomial in ${\bm P}$ of degree
$S$ to find from \re{super-bax1}
\begin{align}\label{super-main}
\Psi_{S,\alpha}({\bm P}) =\xi_{S,\alpha}\, \Phi_{S,\alpha}({\bm Z}) \,,\qquad P_i = Z_i-Z_{i+1}\,,
\end{align}
with $\xi_{S,\alpha} =  i^S / c_Q = i^S/Q_{S,\alpha}(i/2)$. This relation extends the
$SL(2)$ duality symmetry \re{main} to a larger $SL(2|2)$ sector in planar $\mathcal{N}=4$
SYM. As before, the duality relation \re{super-main} involves the polynomial
$\Psi_{S,\alpha}({\bm P})$ evaluated for the total supermomentum equal to zero, $\sum_i
P_i=0$. Nevertheless, similar to the situation in the $SL(2)$ sector, the duality relation
\re{super-main} combined with \re{super-rel} allows us to determine $\Psi_{S,\alpha}({\bm
P})$ for an arbitrary total momentum.

\subsection{Dual superconformal symmetry}

Let us show that the relation between the two sets of variables $P_i=Z_i-Z_{i+1}$ in
\re{super-main} corresponds to the collinear limit of general dual superconformal
transformations \re{dual-coor}. To this end, we consider supersymmetric extension of form
factors introduced in Sect.~2.3
\begin{align}\label{super-ff}
F_{S,\alpha}(\bm{p},\bm{\eta})=\vev{0 | {\mathcal O}_{S,\alpha}(0)  | \bm{p},\bm{\eta}} =\widebar \Psi_{S,\alpha}(\partial_{\bm Z})
\vev{0 |\mathbb{O}({\bm Z}) | \bm{p},\bm{\eta}} \bigg|_{{\bm Z}=0}\,,
\end{align}
where in the second relation we applied \re{O4}. Here $\ket{\bm{p},\bm{\eta}}$ denotes the
on-shell state of $L$ super particles \re{super-state} carrying the light-like momenta
$p_i \bar n^\mu$ and the odd coordinates $\eta_i$.

The super form factor \re{super-ff} has a well-defined expansion in powers of $\eta$'s. The
lowest component of the expansion, $F_{S,\alpha}(\bm{p},0)$ coincides with the scalar form
factor considered in Sect.~2.3 whereas the remaining terms describe form factors evaluated
over the states involving scalars, helicity $(+1/2)$ gaugino and helicity $(+1)$ gluons.
To the leading order in the coupling, we replace each superfield in the light-ray operator \re{super-O} with the
plane wave \re{super-wave} to get
 \begin{align}
 F_{S,\alpha}(\bm{p},\bm{\eta})= L \,
 \tr\lr{T^{a_1}\ldots T^{a_L}}\,\widebar  \Psi_{S,\alpha}(\partial_{\bm Z})
 \prod_{i=1}^L\e^{ip_i z_i +\vev{np_i}(\theta_i\eta_i)  }\bigg|_{{\bm Z}=0}+\text{(perm)}\,,
\end{align}
where `perm' denote terms with permutations of $(p_i,a_i,\eta_i)$ and  we  took into account the 
cyclic symmetry of the polynomial $
\Psi_{S,\alpha}$. In this way, we obtain
\begin{align}\label{super-bare}
  F_{S,\alpha}(x; \bm{p},\bm{\eta})= L \,
 \tr\lr{T^{a_1}\ldots T^{a_L}}\, \Psi_{S,\alpha}({\bm P}) +\text{(perm)},
\end{align}
where the polynomial on the right-hand side is evaluated for  supermomentum $P_k=(ip_k,
\vev{n p_k} \eta_k^a)$.

According to \re{super-main}, the form factor \re{super-bare} is related to the
coefficient function $\Phi_{S,\alpha}({\bm Z})$ when expressed in term of the dual
variables
\begin{align}\label{last}
 p_i  = z_i -z_{i+1}\,,\qquad \vev{n p_i} \eta_i^a = \vartheta_i - \vartheta_{i+1} \,.
\end{align}
Then,   the superconformal $SL(2|2)$ symmetry of  $\Phi_{S,\alpha}({\bm Z})$ is translated
through the duality relation \re{super-main} into the dual superconformal symmetry of the
form factor. Comparing \re{last} with the general expression for the dual variables
\re{dual-coor} we observe that, in the collinear limit \re{config}, the two sets of
variables are related to each other as
\begin{align}
x_i^\mu  = z_i \bar n^\mu  \,,\qquad \theta_i^{\alpha \,a}  =\vartheta_i^a\, { \lambda^\alpha({\bar n})  \over \vev{n \bar n}}\,,
\end{align}
where $\bar n^{\alpha\dot\alpha} = \lambda^\alpha({\bar
n})\tilde\lambda^{\dot\alpha}({\bar n})$. Thus, the dual relation \re{super-main} can be
interpreted as yet another manifestation of the dual superconformal symmetry in planar
$\mathcal N=4$ SYM.

\section{Conclusions and outlook}

In this paper, we have studied the properties of conformal operators in the $SL(2)$ sector
and its supersymmetric extension. The correlation functions of these operators and their
form factors with respect to asymptotic on-shell states are determined in the appropriate
limit by two different polynomials which can be identified as eigenstates of the
dilatation operator in the momentum and coordinate representations. As such, these
polynomials respect the conventional $\mathcal N=4$ superconformal symmetry and are
related to each other through an integral transformation which is analogous to the Fourier
transformation for the discrete series representation of the $SL(2)$ group. We argued that, in virtue of
integrability of the dilatation operator in planar $\mathcal N=4$ SYM, the two polynomials
satisfy a duality relation -- they are proportional to each other upon an appropriate
identification of momenta and coordinates.%
\footnote{The situation here is similar to the well-known property of harmonic oscillator Hamiltonian
  $H=p^2/2+x^2/2$ whose eigenstates in the coordinate and momentum representations are
related to each other through Fourier transform and, at the same time, they coincide (up to an overall
normalisation factor) upon
identification $x\sim p$.}

The duality relation  implies that eigenstates of the $SL(2)$ dilatation operator possess
the $\mathcal N=4$ dual superconformal symmetry. The dilation operator is believed to be
integrable in planar  $\mathcal N=4$ SYM and its eigenvalues can be found for any coupling
\cite{Beisert:2010jr}. The dual conformal symmetry allows us to extend integrability to
the corresponding eigenstates. Indeed we have shown in this paper that the spectrum of
one-loop eigenstates is integrable in the sense that it is uniquely fixed by the dual
symmetry. What happens beyond the leading order? To verify the dual symmetry at higher
loops, we can use the explicit expression for the dilatation operator in the $SL(2|2)$
sector constructed in Ref.~\cite{Zwiebel:2005er}. Going through diagonalisation of this
operator, it can be checked that the dual conformal symmetry survives to two loops
\cite{prep}. All-loop proof of the dual symmetry remains an open problem.

We used the relation between the one-loop dilatation operator and the $SL(2)$ Heisenberg
spin chain to show that the dual symmetry is generated by the Baxter operator
$\mathbb{Q}(u)$, more precisely, by the leading term of its asymptotic expansion for large
values of the spectral parameter $u$. Assuming that the dual conformal symmetry is present
at higher loops, what could be the operator that generalises the Baxter $Q-$operator and
generates the dual symmetry beyond one loop?

We employed the light-cone superfield formalism to obtain supersymmetric extension of the
duality relation. In this formalism various field components in $\mathcal N=4$ SYM can be
combined into a single light-cone super field in such a way that the relations obtained in
the $SL(2)$ sector can be easily extended to the $SL(2|2)$ sector. Trying to extend the
duality relation to larger $SL(2|4)$ sector we encounter the following difficulty. The
$\mathcal N=4$ light-cone superfield has spurious field components (see Appendix~A) whose
contribution to the duality relation should be carefully separated while preserving the
superconformal symmetry.  

Finally, it would be interesting to generalise the duality relation from the $SL(2|4)$
sector to to the full $PSU(2,2|4)$ superconformal group. For this purpose, it is
suggestive to relax the condition for all scalar fields in \re{O-def} to be located along
the same light-ray and consider a more general operator like a supersymmetric light-like
Wilson loop modified by the additional insertions of $\mathcal N=4$ superfields at the
cusp points. Expansion of such operator in powers of like-like distances produces the most
general local Wilson operators in $\mathcal N=4$ SYM. We expect that the duality relation
analogous to \re{super-ann} should hold in this case with the corresponding
super-coordinates related through \re{dual-coor}. This question deserves further investigation.

\section*{Acknowledgements}

We are grateful to A. Belitsky, V. Braun, A. Gorsky and R. Pasechnik for a collaboration on various
topics related to the present work. We would also like to thank V. Bazhanov, E. Sokatchev
and A. Tseytlin for useful discussions. The work was supported in part by the French
National Agency for Research (ANR) under contract StrongInt (BLANC-SIMI-4-2011), by
CNRS/RFFI  (PICS 6076) and CNRS-DNRF grants and by the DFG  grant BR2021/5-2.

\appendix

 \def\theequation{\thesection.\arabic{equation}}

\section{Light-cone superfield and superstates in $\mathcal N=4$ SYM}

In this Appendix, we review how various on-shell states and the corresponding quantum
fields are described in the light-cone superspace approach  in $\mathcal N=4$ SYM.

Asymptotic states in $\mathcal N=4$ SYM include six real scalars, gluons with helicity
$(\pm 1)$ and four gaugino with helicity $(\pm \ft12)$. These states can be combined into
a single  $\mathcal N=4$  superstate by introducing Grassman variables $\eta^A$ (with
$A=1,\dots,4$) \cite{Nair:1988bq}
\begin{align}\notag\label{superstate}
\ket{p,\eta}_{_{\mathcal{N}=4}} = & \big[a_{-1}^\dagger(p)+\eta^A a_{-1/2,A}^{\dagger}(p)
+ \ft12 \eta^A\eta^B \, a_{0,{AB}}^\dagger(p)
\\
& +\ft1{3!} \epsilon_{ABCD}\eta^A\eta^B\eta^C a_{1/2}^{\dagger D }(p)
+\ft1{4!} \epsilon_{ABCD}\eta^A\eta^B\eta^C\eta^D a_{1}^\dagger(p)
 \big] \ket{0}\,,
\end{align}
where $a^\dagger_{h}(p)$ is the creation operator of a particle with helicity $h$ and
light-like momentum $p^\mu$. The scalar and gaugino states carry the $R-$symmetry charge
and their creation operators have the additional $SU(4)$ indices. The odd variables
$\eta^A$ have the helicity $(-1/2)$, so that  each term in the expansion of the superstate
has the same helicity charge $(-1)$.

Each term in the expansion of the superstate \re{superstate} is associated with the
corresponding quantum field. Similar to \re{superstate} we can combine various field
components into a single light-cone superfield by introducing odd coordinates
$\theta_A$ (with $A=1,\dots,4$) \cite{Mandelstam:1982cb,Brink:1982pd,Belitsky:2004sc}
\begin{align}\notag\label{Phi}
\Phi_{_{\mathcal{N}=4}}(x,\theta_A) =& (i n\partial)^{-2} \bar F_n(x) + \theta_A(in\partial)^{-1} \bar \psi_{n}^A(x)  +  \frac12\theta_A\theta_B   \phi^{AB}
\\
& +\ft1{3!}\epsilon^{ABCD} \theta_A\theta_B\theta_C \psi_{n D}+\ft1{4!} \epsilon^{ABCD} \theta_A\theta_B\theta_C\theta_D F_n(x)\,.
\end{align}
Here the coefficients in front of powers of $\theta_A$ involve special components of field
operators which describe independent propagating degrees of freedom in $\mathcal N=4$ SYM
theory quantised on the light-cone cone in the gauge $(n\cdot A(x))=0$ with $n^2=0$. They
are given by gaugino and strength tensor fields projected onto (anti)holomorphic spinors
defining the light-like vector $n^{\alpha\dot\alpha} = \lambda_n^\alpha
\tilde\lambda_n^{\dot\alpha}$
\begin{align}\notag \label{ind-com}
 & \psi^A_n(x) =\lambda^\alpha_n \,\psi_\alpha^A(x)\,, \qquad \quad \bar \psi_{n A}(x) = \tilde \lambda_{n\,\dot\alpha}\bar \psi^{\dot\alpha}_{A}(x)\,,
 \\[2mm]
 & F_{n}(x)=\lambda_n^\alpha \lambda_n^\beta F_{\alpha\beta}(x)
\,, \qquad
\bar F_n(x) = \tilde \lambda_n^{\dot\alpha}\tilde \lambda_n^{\dot\beta} \bar F_{\dot\alpha\dot\beta}(x)\,,
\end{align}
with $F_{\alpha\beta}$ and $\bar F_{\dot\alpha\dot\beta}$ being (anti) self-dual  part of
the strength tensor, $F_{\alpha\dot\alpha,\beta\dot\beta} =
F_{\alpha\beta}\epsilon_{\dot\alpha\dot\beta} + \bar F_{\dot\alpha\dot\beta}
\epsilon_{\alpha\beta}$. The remaining field components can be expressed in terms of those
in \re{ind-com} through the equations of motion.

The operator $\Phi(x,\theta_A)$ creates out of vacuum the $\mathcal N=4$ multiplet of
single particle on-shell states (scalars, gaugino and gluons). Evaluating the matrix
element $\vev{0|\Phi(x,\theta_A)|p,\eta}$ with respect to the superstate \re{superstate}
we find that each term of the expansion \re{Phi} produces a plane wave with the quantum
numbers of the corresponding state
\begin{align}
\vev{0|\Phi(x,\theta)|p,\eta}
= \e^{ipx}\bigg( {1\over \vev{np}^{2}}+{ (\theta\cdot \eta)\over  \vev{np}}+\ft12 (\theta\cdot \eta)^2  +\frac1{3!} \vev{np}(\theta\cdot \eta)^3
+\frac1{4!} \vev{np}^2(\theta\cdot \eta)^4\bigg),
\end{align}
where $(\theta\cdot \eta)=\theta_A \eta^A$ and $\vev{np}=\lambda^\alpha(n)
\lambda_\alpha(p)$. The same relation can be rewritten in a compact form as
\begin{align}\label{pl}
\vev{0|\Phi(x,\theta)|p,\eta}
=\vev{np}^{-2} \e^{ipx+\vev{np}(\theta\cdot \eta)}\,.
\end{align}
It allows us to identify the supermomentum conjugated to the odd coordinate $\theta_A$ to
be $\vev{np}\eta^A$. Notice that the additional factor of $\vev{np}^{-2}$ on the
right-hand side of \re{pl} is needed to match helicity $(-1)$ of the super state \re{superstate}.

The light-cone superfield \re{Phi} belongs to the representation of the $SL(2|4)$
superconformal group of spin $(-1/2)$ \cite{Belitsky:2004sc}. This representation is
reducible   due to the following unusual feature of \re{Phi}. Notice that the
first two terms on the right-hand side of \re{Phi} contain inverse derivatives and,
therefore, the corresponding field operators are nonlocal. Expanding the superfield
\re{Phi}  around $x=0$ we can identify the nonlocal spurious components to be
$\partial_+^{-2} \bar F_n(0)$, $\partial_+^{-1} \bar F_n(0)$ and $\partial_+^{-1} \bar
\psi_{n}^A(0) $ (with $\partial_+\equiv (n\cdot\partial)$) and verify that they define an
irreducible  $SL(2|4)$ component of the $\mathcal N=4$ superfield. The presence of
nonlocal spurious components inside $\Phi(x,\theta)$ complicates the construction of the
dilatation operator in the $SL(2|4)$ sector \cite{Belitsky:2004sc}.

We can avoid the problem with reducibility of $\mathcal N=4$ superfield by using the
formulation of the same gauge theory in terms of $\mathcal N=2$ light-cone superfield. The
latter can be identified as the coefficient in front of $\theta_3\theta_4$ in the
expansion \re{Phi}
\begin{align}
\Phi_{_{\mathcal{N}=2}}(x,\theta_a)=\partial_{\theta_4}\partial_{\theta_3}\Phi_{_{\mathcal{N}=4}}(x,\theta_A)\,.
\end{align}
This superfield depends on two odd coordinates $\theta_a$ (with $a=1,2$) and its explicit
expression is given by \re{superZ} with $Z=\phi^{34}$. It is straightforward to verify
that the spurious components do not appear in $\Phi_{_{\mathcal{N}=2}}(x,\theta_a)$ and,
as a consequence, it belongs to the irreducible representation of the $SL(2|2)$ group of
spin $1/2$. In the similar manner, we can project the $\mathcal N=4$ on-shell superstate
\re{superstate} on its $\mathcal N=4$ component by retaining terms proportional to
$\eta^3\eta^4$
\begin{align}
 \ket{p,\eta^a}_{_{\mathcal{N}=2}} = \partial_{\eta^4}\partial_{\eta^3}\ket{p,\eta^A}_{_{\mathcal{N}=4}}
\end{align}
leading to \re{super-state}.


\begin{thebibliography}{99}

\bibitem{Drummond:2008vq}
  J.~M.~Drummond, J.~Henn, G.~P.~Korchemsky and E.~Sokatchev,
  Nucl.\ Phys.\ B {\bf 828} (2010) 317
  [arXiv:0807.1095 [hep-th]].

\bibitem{Beisert:2010jr}
  N.~Beisert, C.~Ahn, L.~F.~Alday, Z.~Bajnok, J.~M.~Drummond, L.~Freyhult, N.~Gromov and R.~A.~Janik {\it et al.},
  Lett.\ Math.\ Phys.\  {\bf 99} (2012) 3
  [arXiv:1012.3982 [hep-th]].

\bibitem{Alday:2007hr}
  L.~F.~Alday and J.~M.~Maldacena,
  JHEP {\bf 0706} (2007) 064
  [arXiv:0705.0303 [hep-th]].

\bibitem{Drummond:2007aua}
  J.~M.~Drummond, G.~P.~Korchemsky and E.~Sokatchev,
  Nucl.\ Phys.\ B {\bf 795} (2008) 385
  [arXiv:0707.0243 [hep-th]].

\bibitem{Brandhuber:2007yx}
  A.~Brandhuber, P.~Heslop and G.~Travaglini,
  Nucl.\ Phys.\ B {\bf 794} (2008) 231
  [arXiv:0707.1153 [hep-th]].

 \bibitem{Mason:2010yk}
  L.~J.~Mason and D.~Skinner,
  JHEP {\bf 1012} (2010) 018
  [arXiv:1009.2225 [hep-th]].

 \bibitem{CaronHuot:2010ek}
  S.~Caron-Huot,
  JHEP {\bf 1107} (2011) 058
  [arXiv:1010.1167 [hep-th]].

\bibitem{Belitsky:2011zm}
  A.~V.~Belitsky, G.~P.~Korchemsky and E.~Sokatchev,
  Nucl.\ Phys.\ B {\bf 855} (2012) 333
  [arXiv:1103.3008 [hep-th]].

\bibitem{Eden:2011yp}
  B.~Eden, P.~Heslop, G.~P.~Korchemsky and E.~Sokatchev,
  Nucl.\ Phys.\ B {\bf 869} (2013) 329
  [arXiv:1103.3714 [hep-th]].

\bibitem{Eden:2011ku}
  B.~Eden, P.~Heslop, G.~P.~Korchemsky and E.~Sokatchev,
  Nucl.\ Phys.\ B {\bf 869} (2013) 378
  [arXiv:1103.4353 [hep-th]].

\bibitem{Adamo:2011dq}
  T.~Adamo, M.~Bullimore, L.~Mason and D.~Skinner,
  JHEP {\bf 1108} (2011) 076
  [arXiv:1103.4119 [hep-th]].

\bibitem{Broadhurst:1993ib}
  D.~J.~Broadhurst,
  Phys.\ Lett.\ B {\bf 307} (1993) 132.

\bibitem{Drummond:2006rz}
  J.~M.~Drummond, J.~Henn, V.~A.~Smirnov and E.~Sokatchev,
  JHEP {\bf 0701} (2007) 064
  [hep-th/0607160].

\bibitem{Lipatov:1998as}
  L.~N.~Lipatov,
  Nucl.\ Phys.\ B {\bf 548} (1999) 328
  [hep-ph/9812336].

\bibitem{Gomez:2009bx}
  C.~Gomez, J.~Gunnesson and A.~S.~Vera,
  Phys.\ Lett.\ B {\bf 690} (2010) 78
  [arXiv:0908.2568 [hep-th]].

\bibitem{Prygarin:2009zz}
  A.~Prygarin,
  Phys.\ Rev.\ C {\bf 83} (2011) 055206
  [arXiv:0911.5279 [hep-ph]].

\bibitem{Braun:1999te}
  V.~M.~Braun, S.~E.~Derkachov, G.~P.~Korchemsky and A.~N.~Manashov,
  Nucl.\ Phys.\ B {\bf 553} (1999) 355
  [hep-ph/9902375].

\bibitem{Nair:1988bq}
  V.~P.~Nair,
  Phys.\ Lett.\ B {\bf 214} (1988) 215.

\bibitem{Drummond:2007au}
  J.~M.~Drummond, J.~Henn, G.~P.~Korchemsky and E.~Sokatchev,
  Nucl.\ Phys.\ B {\bf 826} (2010) 337
  [arXiv:0712.1223 [hep-th]].

\bibitem{Korchemsky:2009hm}
  G.~P.~Korchemsky and E.~Sokatchev,
  Nucl.\ Phys.\ B {\bf 832} (2010) 1
  [arXiv:0906.1737 [hep-th]].

\bibitem{CaronHuot:2011kk}
  S.~Caron-Huot and S.~He,
  JHEP {\bf 1207} (2012) 174
  [arXiv:1112.1060 [hep-th]].

\bibitem{Bullimore:2011kg}
  M.~Bullimore and D.~Skinner,
  arXiv:1112.1056 [hep-th].

\bibitem{Kallosh:1998ji}
  R.~Kallosh and A.~A.~Tseytlin,
  JHEP {\bf 9810} (1998) 016
  [hep-th/9808088].

\bibitem{Berkovits:2008ic}
  N.~Berkovits and J.~Maldacena,
  JHEP {\bf 0809} (2008) 062
  [arXiv:0807.3196 [hep-th]].

\bibitem{Beisert:2008iq}
  N.~Beisert, R.~Ricci, A.~A.~Tseytlin and M.~Wolf,
  Phys.\ Rev.\ D {\bf 78} (2008) 126004
  [arXiv:0807.3228 [hep-th]].

\bibitem{Mandal:2002fs}
  G.~Mandal, N.~V.~Suryanarayana and S.~R.~Wadia,
  Phys.\ Lett.\ B {\bf 543} (2002) 81
  [hep-th/0206103].

\bibitem{Bena:2003wd}
  I.~Bena, J.~Polchinski and R.~Roiban,
  Phys.\ Rev.\ D {\bf 69} (2004) 046002
  [hep-th/0305116].

\bibitem{Lepage:1980fj}
  G.~P.~Lepage and S.~J.~Brodsky,
  Phys.\ Rev.\ D {\bf 22} (1980) 2157.
 
\bibitem{Chernyak:1983ej}
  V.~L.~Chernyak and A.~R.~Zhitnitsky,
  Phys.\ Rept.\  {\bf 112} (1984) 173.
  
\bibitem{Braun:2003rp}
  V.~M.~Braun, G.~P.~Korchemsky and D.~Mueller,
  Prog.\ Part.\ Nucl.\ Phys.\  {\bf 51} (2003) 311
  [hep-ph/0306057].
 
\bibitem{Baxter:1972hz}
  R.~J.~Baxter,
  Annals Phys.\  {\bf 70} (1972) 193
   [Annals Phys.\  {\bf 281} (2000) 187].

\bibitem{Bazhanov:1989nc}
  V.~V.~Bazhanov and Y.~.G.~Stroganov,
  J.\ Statist.\ Phys.\  {\bf 59} (1990) 799.

\bibitem{Gaudin:1992ci}
  M.~Gaudin and V.~Pasquier,
  J.\ Phys.\ A {\bf 25} (1992) 5243.

\bibitem{Bazhanov:1996dr}
  V.~V.~Bazhanov, S.~L.~Lukyanov and A.~B.~Zamolodchikov,
  Commun.\ Math.\ Phys.\  {\bf 190} (1997) 247
  [hep-th/9604044].


\bibitem{Derkachov:1999pz}
  S.~E.~Derkachov,
  J.\ Phys.\ A {\bf 32} (1999) 5299
  [solv-int/9902015].

\bibitem{Brandhuber:2010ad}
  A.~Brandhuber, B.~Spence, G.~Travaglini and G.~Yang,
  JHEP {\bf 1101} (2011) 134
  [arXiv:1011.1899 [hep-th]].

\bibitem{Alday:2007he}
  L.~F.~Alday and J.~Maldacena,
  JHEP {\bf 0711} (2007) 068
  [arXiv:0710.1060 [hep-th]].

\bibitem{Mandelstam:1982cb}
  S.~Mandelstam,
  Nucl.\ Phys.\ B {\bf 213} (1983) 149.

\bibitem{Brink:1982pd}
  L.~Brink, O.~Lindgren and B.~E.~W.~Nilsson,
  Nucl.\ Phys.\ B {\bf 212} (1983) 401.

\bibitem{Belitsky:2004sc}
  A.~V.~Belitsky, S.~E.~Derkachov, G.~P.~Korchemsky and A.~N.~Manashov,
  Nucl.\ Phys.\ B {\bf 708} (2005) 115
  [hep-th/0409120].

\bibitem{Belitsky:2005gr}
  A.~V.~Belitsky, S.~E.~Derkachov, G.~P.~Korchemsky and A.~N.~Manashov,
  Nucl.\ Phys.\ B {\bf 722} (2005) 191
  [hep-th/0503137].


\bibitem{Belitsky:2006cp}
  A.~V.~Belitsky, S.~E.~Derkachov, G.~P.~Korchemsky and A.~N.~Manashov,
  J.\ Stat.\ Mech.\  {\bf 0701} (2007) P01005
  [hep-th/0610332].

\bibitem{Bazhanov:2008yc}
  V.~V.~Bazhanov and Z.~Tsuboi,
  Nucl.\ Phys.\ B {\bf 805} (2008) 451
  [arXiv:0805.4274 [hep-th]].

\bibitem{prep} G.~P.~Korchemsky,  to appear.

\bibitem{Derkachov:1997qv}
  S.~E.~Derkachov, S.~K.~Kehrein and A.~N.~Manashov,
  Nucl.\ Phys.\ B {\bf 493} (1997) 660.

\bibitem{Makeenko:1980bh}
  Y.~M.~Makeenko,
  Sov.\ J.\ Nucl.\ Phys.\  {\bf 33} (1981) 440
   [Yad.\ Fiz.\  {\bf 33} (1981) 842].

\bibitem{Ohrndorf:1981qv}
  T.~Ohrndorf,
  Nucl.\ Phys.\ B {\bf 198} (1982) 26.

\bibitem{Balitsky:1987bk}
  I.~I.~Balitsky and V.~M.~Braun,
  Nucl.\ Phys.\ B {\bf 311} (1989) 541.

\bibitem{Braun:1998id}
  V.~M.~Braun, S.~E.~Derkachov and A.~N.~Manashov,
  Phys.\ Rev.\ Lett.\  {\bf 81} (1998) 2020
  [hep-ph/9805225].

 \bibitem{Belitsky:1999qh}
  A.~V.~Belitsky,
  Phys.\ Lett.\ B {\bf 453} (1999) 59
  [hep-ph/9902361].


\bibitem{FST} L.~D.~Faddeev, E.~K.~Sklyanin and L.~A.~Takhtajan,
  Theor.\ Math.\ Phys.\  {\bf 40} (1980) 688
   [Teor.\ Mat.\ Fiz.\  {\bf 40} (1979) 194].

\bibitem{KS} P.~P.~Kulish and E.~K.~Sklyanin,
  Lect.\ Notes Phys.\  {\bf 151} (1982) 61.

\bibitem{Skl} E.~K.~Sklyanin, {\it Quantum Inverse Scattering Method.Selected Topics}, in
    "Quantum Group and Quantum Integrable Systems" (Nankai Lectures in Mathematical
    Physics), ed. Mo-Lin Ge, Singapore: World Scientific, 1992, pp.63-97; hep-th/9211111.

\bibitem{Fad} L.~D.~Faddeev,~{\it How Algebraic Bethe Anstz works for integrable model},
    In: Quantum Symmetries/Symetries Qantiques, Proc.Les-Houches summer school, LXIV. Eds.
    A.Connes, K.Kawedzki, J.Zinn-Justin. North-Holland, 1998, 149-211, hep-th/9605187.

\bibitem{Zwiebel:2005er}
  B.~I.~Zwiebel,
  JHEP {\bf 0602} (2006) 055
  [hep-th/0511109].

\end{thebibliography}
\end{document}